\documentclass[a4paper,11pt]{article}

\usepackage{jcappub} 
\usepackage{color}
\usepackage{graphicx,amsmath,dcolumn,bm}
\usepackage{hyperref}
\usepackage{epstopdf}

\title{\boldmath Testing consistency of general relativity with kinematic and dynamical probes}

\author[a]{Xiao-Wei Duan,}
\author[a]{Min Zhou,}
\author[a]{Tong-Jie Zhang}

\affiliation[a]{Department of Astronomy, Beijing Normal University,\\Beijing 100875, P.R.China}

\emailAdd{duanxw@mail.bnu.edu.cn}
\emailAdd{tjzhang@bnu.edu.cn}

\abstract{In this work, we test consistency relations between a kinematic probe, the observational Hubble data, and a dynamical probe, the growth rates for cosmic large scale structure, which should hold if general relativity is the correct theory of gravity on cosmological scales. Moreover, we summarize the development history of parametrization in testings and make an improvement of it. Taking advantage of the Hubble parameter given from both parametric and non-parametric methods, we propose three equations and test two of them performed by means of two-dimensional parameterizations, including one using trigonometric functions we propose.
As a result, it is found that the consistency relations satisfies well at $1\sigma$ CL and trigonometric functions turn out to be efficient tools in parameterizations. Furthermore, in order to confirm the validity of our test, we introduce a model of modified gravity, DGP model and compare the testing results in the cases of $\Lambda$CDM, ``DGP in GR'' and DGP model with mock data. It can be seen that it is the establishing of consistency relations which dominates the results of the testing. Overall, the present observational Hubble data and growth rate data favor convincingly that the general relativity is the correct theory of gravity on cosmological scales.}

\begin{document}
\maketitle
\flushbottom

\section{Introduction}
\label{sec:intro}

Einstein's theory of general relativity (GR) has, unchangeably,  remained evergreen
at the heart of astrophysics over almost one century after its formulation.
The testing of the general relativity theory takes the central position all the time in the modern physics\cite{Berti2015}.
The theory has already passed
the precisive experimental tests to data on the scale of the Solar System and smaller ones with
flying colors. Naturally, testing general relativity on cosmological scales\cite{Peebles2004}
will be the current and future target for gravitational physics.

The cosmological observations, testing the general relativity,  include two traditional classes of probes\cite{Perivolaropoulos2010}.
One class is the so-called "geometric probes", which includes Type Ia supernovae (as standard candles), Baryon Acoustic Oscillations (BAO)
and geometric properties of weak lensing. These probes can determinate the Hubble parameter $H(z)$
as a function of the redshift $z$ through angular or luminosity distances. As the ramification of geometric probes, we call $H(z)$ a "kinematic probe"  because of its kinematic origin.
The Hubble parameter $H(z)$
is defined as $H = \dot{a}/a$, where $a$ denotes the cosmic scale factor and $\dot{a}$
is its rate of change with respect to the cosmic time (the age of the universe when the observed photon is emitted).
Moreover, the cosmic scale factor $a$ is related to the redshift $z$ by the formula $a(t )/a(t_0) = 1/(1+z)$, where $t_0$ denotes the
current time which is taken to be a constant. The observational $H(z)$ data (OHD) are directly
related to the expansion history of the universe. The other class is the so-called "dynamical probes",
including weak lensing, galaxy clustering  and  redshift-space distortions. The dynamical probes
can measure the gravitational law on cosmological scales, main of which is the evolution of
linear density perturbations $\delta(z)$, where $\delta(z) = (\rho - \bar{\rho})/\rho$ is the
overdensity. In order to obtain observable data, the growth rate can be defined as a combination of the structure growth,  $f(z) = dln\delta/dlna$, and the redshift-dependent rms fluctuations of the linear density field, $\sigma_8(z)$, $R(z)(f\sigma_8(z))\equiv f(z)\sigma_8(z)$. For the cosmic large scale structure, both of them can be measured from cosmological data. Therefore, the kinematic probes and dynamical probes provide complementary measurements of the nature of the observed cosmic acceleration\cite{Lue2004,Lue2006,Heavens2007,Zhang2007}.


In this paper, we will investigate consistency relations between the kinematic and dynamical probes in the framework
of general relativity. Our work is partly motivated by the consistency tests in cosmology proposed
by Knox et al.\cite{Knox2006}, Ishak et al.\cite{Ishak2006}, T. Chiba and R. Takahashi\cite{Chiba2007} and Yun Wang\cite{YW2008}.
The consistency relation from T. Chiba and R. Takahashi, which is constructed theoretically between the luminosity distance and the density perturbation, relies on the matter density parameter today, and it is hard to achieve using available observed data. Linder et al. \cite{Linder2005} introduced the gravitational growth index $\gamma$ which is defined by $d\rm{ln}(\delta/a)/d\rm{ln}a=\Omega_m(a)^\gamma-1$ to finish the model-testing\cite{Huterer2007,Linder2007,Pouri2013}. It was initially considered not very insensitive to the equation of state of dark energy\cite{Chiba2007}, but the measurement of $\gamma$ shows us that it can be used to distinguish between different models and it is clear that the growth of matter perturbations can be an efficient dynamical probe to show us the nature of dark energy\cite{Polarski2008,Gannouji2008}. Inspired by these works, we wonder whether we can step a little further. Is there a more direct way to set the test? We understand that the most accurate information about the cosmic expansion rate, $H(z)$,
can provide the unique prediction for the growth rate data.
The obtained $data_{cal}(z)$ must be consistent with the observed $data_{obs}(z)$ if general relativity is the correct theory of gravity in the universe. Given that the assumptions used in the methods to obtain observed data may make specific cosmological models involved in, we choose $f\sigma_8$ for test implementing, which are produced using almost model-independent way. Using the kinematic probe $H(z)$, our testing methods are independent of any cosmological constant from theoretical calculation according to general relativity. The presence of significant deviations from the consistency can be used as the signature of a breakdown of general relativity on
cosmological scales. It is hoped to gain new knowledge about testing the general relativity.

This paper is organized as follows. In Section II we briefly describe the observable data and the relation between Hubble parameter and growth rate, particularly how to establish the consistency relation equations. The analysis methods, calculated results and discussion are presented in Section III. Furthermore, in Section IV, we try to check the validity of this test. And finally, the summary is given in Section V.

\vskip 0.5cm

\section{Observable data  and  relation
between  Hubble parameter  and  growth rate}

The kinematic probe $H(z)$ directly measure the cosmic metric while dynamical probes not only measure the cosmic metric but also the gravitational law concurrently on cosmological scales. Converging these two kinds of probes indicates that our universe is accelerating in its expansion. In our work, we choose $H(z)$, the kinematic probe and the growth rate deduced from density fluctuation as the dynamical one to establish the consistency equation.

\subsection{Kinematic Probes: Observational $H(z)$ Data}
The charming potential of OHD to constrain cosmological parameters and distinguish different
models has been surfacing over in recent years. $H(z)$ can be produced by model-independent direct
observations, and three methods have been developed to measure them up to now \cite{Zhang2010}: galaxy differential age, radial BAO size and gravitational-wave standard sirens methods. In practice, $H(z)$ is usually utilized as a function of the redshift $z$, getting its expression as:
\begin{eqnarray}
 H(z)=-\frac{1}{1+z}\frac{dz}{dt}.
\end{eqnarray}

According this expression, $H(z)$ got its initial measuring method: differential age method.
It was firstly put forward by Jimenez \& Loeb \cite{Jimenez2002} in 2002 using relative galaxy ages, demonstrated by Jimenez et al.\cite{Jimenez2003} with the first $H(z)$ measurement at $z\approx1$ reported. Soon afterwards, Simon et al.\cite{Simon2005} provided eight additional $H(z)$ data points in the redshift value range from 0.17 to 1.75 from the relative ages of passively evolving galaxies (henceforward ¡°SVJ05¡±) and use them to constrain the redshift dependency of DE potential. In 2010, Stern et al. \cite{Stern2010} expanded the dataset (henceforward ¡°SJVKS10¡±) with two new determinations and constrain DE parameters and the spatial curvature via combining the dataset with CMB data. Moreover, Moresco et al. \cite{Moresco2012} utilized the differential spectroscopic evolution of early-type, massive, red elliptical galaxies (can be used as standard cosmic chronometers) and got eight new data points. As time goes by, bigger and better instruments are used to make cosmological observations. Chuang \& Wang \cite{ChuangWang2012} measured $H(z)$ at $z=0.35$ with luminous red galaxies from Sloan Digital Sky Survey (SDSS) Data Release 7 (DR7). Also with applying the galaxy differential age method to SDSS DR7, Zhang et al. \cite{Zhang2014} obtained four new measurements. Moresco \cite{Moresco2015} reported two latest data points with near-infrared spectroscopy of high redshift galaxies. And, recently, they has enriched the dataset again with 5 points through the Baryon Oscillation Spectroscopic Survey (BOSS) Data Release 9, which yield a 6\% accuracy constraint of $H(z=0.4293)=91.8\pm5.3kms^{-1}Mpc^{-1}$\cite{Moresco2016}.

Apart from the differential method, there is another method to obtain $H(z)$ : detection of radial BAO features. It was first utilized by Gazta$\tilde{\rm{n}}$aga et al. \cite{Gaztanaga2009} to get two new $H(z)$ data points in 2009. They used the BAO peak position as a standard ruler in the radial direction. From then on, the dataset on $H(z)$ is expanded: 3 data points from Blake et al. \cite{Blake2012} by combining the measurements of BAO peaks and the Alcock-Paczynski distortion; one data point from Samushia et al. \cite{Samushia2013} using the BOSS DR9 CMASS sample; one data point from Xu et al. \cite{Xu2013} by means of BAO signals from the SDSS DR7 luminous red galaxy sample. Busca et al. \cite{Busca2013}, Font-Ribera et al. \cite{Font-Ribera2014} and Delubac et al. \cite{Delubac2015} provided respectively one data point on $H(z)$ by BAO features in the Lyman-$\alpha$forest of SDSS-III quasars.

Further more, the detection of gravitational wave also provides a way to get $H(z)$: gravitational-wave standard sirens. With future gravitational wave detectors, it will be able to measure source luminosity distances, as integrated quantities of $H^{-1}(z)$, previously out to $z\sim5$ \cite{Bonvin2006,Nishizawa2010}. The quantity of the available data will promisingly be improved dramatically in the near future. As for this work, we put all the available $H(z)$ data up to now into use. After getting rid of overlapped data points with same sources, table \ref{tab:Hubble} lists the precise numerical values of $H(z)$ data points with the corresponding errors, which is collected by Meng et al. \cite{Meng2015} and added with \cite{Moresco2016}. The correspondence of the columns is as follows: redshift, observed Hubble parameter ($km s^{-1} Mpc^{-1}$), used method (I: differential age method, II: BAO method) and references.

\begin{table}[tbp]
\centering
\begin{tabular}{|lccc|}
\hline
{$z$}   & $H(z)$ & Method & Ref.\\
\hline
$0.0708$   &  $69.0\pm19.68$      &  I    &  Zhang et al. (2014)-\cite{Zhang2014}   \\
    $0.09$       &  $69.0\pm12.0$        &  I    &  Jimenez et al. (2003)-\cite{Jimenez2003}   \\
    $0.12$       &  $68.6\pm26.2$        &  I    &  Zhang et al. (2014)-\cite{Zhang2014}  \\
    $0.17$       &  $83.0\pm8.0$          &  I    &  Simon et al. (2005)-\cite{Simon2005}     \\
    $0.179$     &  $75.0\pm4.0$          &  I    &  Moresco et al. (2012)-\cite{Moresco2012}     \\
    $0.199$     &  $75.0\pm5.0$          &  I    &  Moresco et al. (2012)-\cite{Moresco2012}     \\
    $0.20$         &  $72.9\pm29.6$        &  I    &  Zhang et al. (2014)-\cite{Zhang2014}   \\
    $0.240$     &  $79.69\pm2.65$      &  II   &  Gazta$\tilde{\rm{n}}$aga et al. (2009)-\cite{Gaztanaga2009}   \\
    $0.27$       &  $77.0\pm14.0$        &  I    &    Simon et al. (2005)-\cite{Simon2005}   \\
    $0.28$       &  $88.8\pm36.6$        &  I    &  Zhang et al. (2014)-\cite{Zhang2014}   \\
    $0.35$       &  $84.4\pm7.0$          &  II   &   Xu et al. (2013)-\cite{Xu2013}  \\
    $0.352$     &  $83.0\pm14.0$        &  I    &  Moresco et al. (2012)-\cite{Moresco2012}   \\
    $0.3802$     &  $83.0\pm13.5$        &  I    &  Moresco et al. (2016)-\cite{Moresco2016}   \\
    $0.4$         &  $95\pm17.0$           &  I    &  Simon et al. (2005)-\cite{Simon2005}     \\
    $0.4004$     &  $77.0\pm10.2$        &  I    &  Moresco et al. (2016)-\cite{Moresco2016}   \\
    $0.4247$     &  $87.1\pm11.2$        &  I    &  Moresco et al. (2016)-\cite{Moresco2016}   \\
    $0.43$     &  $86.45\pm3.68$        &  II   &  Gaztanaga et al. (2009)-\cite{Gaztanaga2009}   \\
    $0.44$       & $82.6\pm7.8$           &  II   &  Blake et al. (2012)-\cite{Blake2012}  \\
    $0.4497$     &  $92.8\pm12.9$        &  I    &  Moresco et al. (2016)-\cite{Moresco2016}   \\
    $0.4783$     &  $80.9\pm9.0$        &  I    &  Moresco et al. (2016)-\cite{Moresco2016}   \\
    $0.48$       &  $97.0\pm62.0$        &  I    &  Stern et al. (2010)-\cite{Stern2010}     \\
    $0.57$       &  $92.4\pm4.5$          &  II   &  Samushia et al. (2013)-\cite{Samushia2013}   \\
    $0.593$     &  $104.0\pm13.0$      &  I    &  Moresco et al. (2012)-\cite{Moresco2012}   \\
    $0.6$         &  $87.9\pm6.1$          &  II   &  Blake et al. (2012)-\cite{Blake2012}   \\
    $0.68$       &  $92.0\pm8.0$          &  I    &  Moresco et al. (2012)-\cite{Moresco2012}   \\
    $0.73$       &  $97.3\pm7.0$          &  II   &  Blake et al. (2012)-\cite{Blake2012}  \\
    $0.781$     &  $105.0\pm12.0$      &  I    &  Moresco et al. (2012)-\cite{Moresco2012}   \\
    $0.875$     &  $125.0\pm17.0$      &  I    &  Moresco et al. (2012)-\cite{Moresco2012}   \\
    $0.88$       &  $90.0\pm40.0$        &  I    &  Stern et al. (2010)-\cite{Stern2010}     \\
    $0.9$         &  $117.0\pm23.0$      &  I    &  Simon et al. (2005)-\cite{Simon2005}  \\
    $1.037$     &  $154.0\pm20.0$      &  I    &  Moresco et al. (2012)-\cite{Moresco2012}   \\
    $1.3$         &  $168.0\pm17.0$      &  I    &  Simon et al. (2005)-\cite{Simon2005}     \\
    $1.363$     &  $160.0\pm33.6$      &  I    &  Moresco (2015)-\cite{Moresco2015}  \\
    $1.43$       &  $177.0\pm18.0$      &  I    &  Simon et al. (2005)-\cite{Simon2005}     \\
    $1.53$       &  $140.0\pm14.0$      &  I    &  Simon et al. (2005)-\cite{Simon2005}     \\
    $1.75$       &  $202.0\pm40.0$      &  I    &  Simon et al. (2005)-\cite{Simon2005}     \\
    $1.965$     &  $186.5\pm50.4$      &  I    &   Moresco (2015)-\cite{Moresco2015}  \\
    $2.34$       &  $222.0\pm7.0$        &  II   &  Delubac et al. (2015)-\cite{Delubac2015}   \\
\hline
\end{tabular}
\caption{\label{tab:Hubble} The currently available OHD dataset.}
\end{table}

In order to set up a model-independent consistency relation to test general relativity, firstly we need to obtain a good expression on $H(z)$. From the Friedmann equation, $H(z)$ can be written as
\begin{eqnarray}\label{eq:modeleq}
  \frac{H^2}{H^2_0} & = & \Omega_{m_0}(1+z)^3+\Omega_{r_0}(1+z)^4+\Omega_{k_0}(1+z)^2  \nonumber \\
                   &   & +\Omega_{x}{\rm{exp}}[3\int^z_0(1+\omega_x(z))d{\rm{ln}}(1+z)].
\end{eqnarray}
Moreover, when $\omega=-1$ (vacuum energy), it reduces to
\begin{eqnarray}\label{eq:modeleq1}
 \frac{H^2}{H^2_0} & = & \Omega_{m_0}(1+z)^3+\Omega_{r_0}(1+z)^4  \nonumber \\
                   &   & +\Omega_{k_0}(1+z)^2+\Omega_{\Lambda_0}.
\end{eqnarray}
For lower redshift and flat model, it further reduces to
\begin{eqnarray}\label{eq:modeleq2}
 \frac{H^2}{H^2_0} & = & \Omega_{m_0}(1+z)^3+\Omega_{\Lambda_0}.
\end{eqnarray}
It occurred to us that $H^2$ can be parameterized. If we use 4th degree polynomial to fit $H^2$, we can see that this expression can directly degenerate into Eq.~\ref{eq:modeleq1} and this method is a very general way of data modeling. But based on the current observed data, the fitting result of it has a negative coefficient for fourth power term. It will lead  fitting curve of $H^2$ to be negative when the redshift getting bigger, which is unreasonable.

Polynomial fitting suffers from the imprecision of current observed data. So we adopted a general choice:
\begin{eqnarray}\label{eq:Hz}
H^2(z)=a(1+z)^b+c,
\end{eqnarray}
which set exponent into free. It may give more favor to $\Lambda$CDM model but it keeps a more convincing trace when fitting current data. Therefore, we set it as the parametric method (Method A) in our test. Visualization of our fitting curve is shown in Figure 1 (the blue solid line) and the error range is derived from simultaneous predicting method \cite{Matlabd1}.

\begin{figure}[tbp]
\centering 
\includegraphics[width=90.3mm,height=7cm]{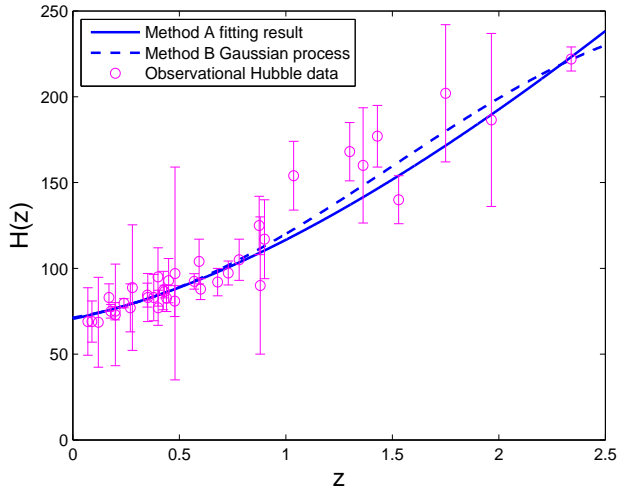}
\caption{\label{fig:H1} Visualization of $\chi^2$ approach to the observational $H(z)$ dataset (Method A) and model-independent reconstruction of it using Gaussian processes (Method B). The solid curve represents the best fitting result in method A and the dashed line refers to the reconstruction with GaPP in method B. The points with error bars denote OHD.}
\end{figure}

\begin{figure}[tbp]
\centering 
\includegraphics[width=90.3mm,height=7cm]{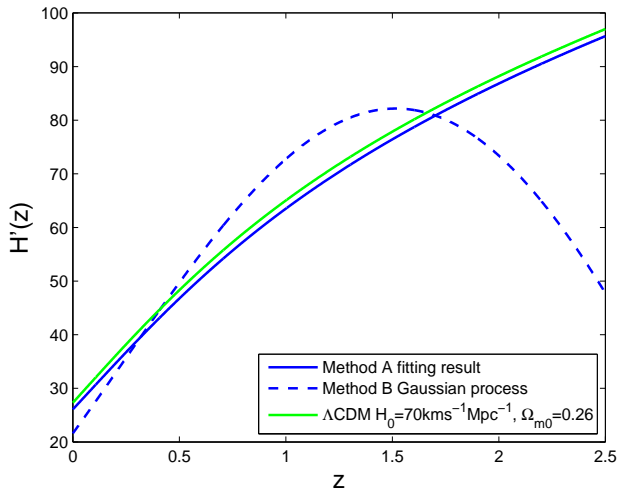}
\caption{\label{fig:H2} Visualization of model-independent reconstruction of the first derivative of $H(z)$ using Method A and B. The solid curve represents result in method A and the dashed line refers to the reconstruction with GaPP in method B. Here is also $H(z)$ reconstruction in $\Lambda$CDM plotted with green line for reference.}
\end{figure}

As for our second method (Method B), we use Gaussian Processes (GP), which is a non-parametric method for smoothing the observational data. The advantage of this method is that it is  also model-independent as Method A and it can perform a reconstruction
of a function from data without assuming a parametrization of the function.  A Gaussian process is the generalization of a Gaussian distribution on a random variable. It describes a distribution over functions. The freedom in Gaussian process comes in the choice of covariance function $k(z,\tilde{z})$,  which determines how smooth the process is and how nearby points such as $z$ and $\tilde{z}$ are correlated.  The covariance function $k(z,\tilde{z})$  depends on a set of hyperparameters $\ell$ and $\sigma_f$, which can be determined by Gaussian process with the observed data.  The detailed analysis and descriptions of Gaussian process method can be found in \cite{Seikel2012}, \cite{Seikel20121} and \cite{Seikel2013}.  Here we use the Gaussian processes in Python (GaPP) \cite{Seikel2012} to reconstruct the Hubble parameter as a function of the redshift from the observational Hubble data.

We use 38 $H(z)$ measurements to obtain the model-independent reconstruction of $H(z)$ and the first derivative of it using Gaussian processes. The redshift range of observational $H(z)$ data is [0.0708,2.36] so that we set $z$ variable as [0,2.5]. Our results for the non-parametric approach are shown in Figure \ref{fig:H1} and \ref{fig:H2}. Figure 1 also gives the observational $H(z)$ data with their respective errorbars and makes the simulated $H(z)$ data obtained from the two methods we use compared with each other. The blue solid line refers to the fitting result in Method A, with the dash line denoting production from Method B. It is obvious in Figure 1 that method A and B capture data correctly though there is a slight difference between them. The reconstructions of first derivative of $H(z)$ are displayed in Figure 2. There is also $H(z)$ reconstruction in $\Lambda$CDM plotted with green line for reference. It is obvious that confined by the quality of current data, the trend of $dH(z)/dz$ is not very reasonable. Nevertheless, Gaussian Processes have the limitation that they can't keep reconstructing data exactly all the time as the argument getting bigger.

\subsection{Dynamical Probes: the Growth Rate deduced from Density Fluctuation}

Through the observations of distant supernovae, we know that the Universe is in a phase of accelerated expansion now. Different models, though producing similar expansion rates, are corresponding to different growth rate of large-scale structure with cosmic time \cite{Peebles1980}. The growth of cosmic structure is driven by the motion of matter, while galaxies act as ``tracer particles''. Coherent galaxy motions, reconstructed by galaxy redshift surveys, can introduce a radial anisotropy in the clustering pattern \cite{LMBS2008}. Moreover, in linear perturbation theory it is possible to establish an expression to describe the growth of a generic small-amplitude density fluctuation $\delta(z)$, which is defined as $\delta_m\equiv\delta\rho_m/\rho_m$, by general relativity. It is expressed as \cite{LGuzzo2008]}:
\begin{eqnarray}\label{eq:defidensitydelta}
\ddot{ \delta}+2H\dot{\delta}-4{\pi}G\rho_M{\delta} &=& 0,
\end{eqnarray}
where a dot represents the derivative with respect to the cosmic time $t$ and $\rho_M$ denotes the matter density. The linear growth rate $f(z)$ can be defined as
\begin{eqnarray}\label{eq:fdefinition}
f(z)\equiv\frac{dln\delta}{dlna},
\end{eqnarray}
to measure how rapidly the structure is being assembled according to different redshifts. However, most of the observed data on $f(z)$ we obtained up to now let specific cosmological models (mostly $\Lambda$CDM) play quite big role in the process to produce them. So we turned our eyes toward another growth rate dataset: $R(z)$, which is almost a model-independent way to express the observed growth history of the universe \cite{Song2009}. It is combined by the growth rate of structure $f(z)$ and the redshift-dependent rms fluctuations of the linear density field $\sigma_8$ as
\begin{eqnarray}\label{eq:fsigma8definition}
R(z)(f\sigma_8(z))\equiv f(z)\sigma_8(z),
\end{eqnarray}
where $\sigma_8(z)=\sigma_8(0)\frac{\delta(z)}{\delta(0)}$. The used dataset is shown in Table \ref{tab:R},
which summarizes the used numerical values of the observational data on $R(z)$ with the corresponding errors. The correspondence of the columns in it is as follows: redshift, observed growth rate (henceforward $R_{obs}(z)$), references.

\begin{table}[tbp]
\centering
\begin{tabular}{|lcc|}
\hline
{$z$}   & $R_{obs}(z)$  & Ref. \\
\hline
$0.02$   &  $0.360\pm0.040$          &  \cite{Hudson2012}   \\
    $0.067$       &  $0.423\pm0.055$            &  \cite{Beutler2012}   \\
    $0.17$     &  $0.510\pm0.060$             &  \cite{Song2009}     \\
    $0.25$     &  $0.3512\pm0.0583$         &  \cite{Samushia2012}   \\
    $0.30$       &  $0.407\pm0.055$          &   \cite{Tojeiro2012}   \\
    $0.32$    &  $0.384\pm0.095$       &  \cite{Chuang201312}   \\
    $0.35$       &  $0.440\pm0.050$           &  \cite{Song2009} \\
    $0.35$       &  $0.445\pm0.097$         &   \cite{Chuang2013}  \\
    $0.37$     &  $0.4602\pm0.0378$            &  \cite{Samushia2012}   \\
    $0.40$      &  $0.419\pm0.041$            &  \cite{Tojeiro2012}    \\
    $0.44$       &  $0.413\pm0.080$            &  \cite{Blake2012}     \\
    $0.50$       &  $0.427\pm0.043$        &  \cite{Tojeiro2012}   \\
    $0.57$     &  $0.423\pm0.052$         &  \cite{Samushia2014}   \\
    $0.60$         &  $0.433\pm0.067$            &  \cite{Tojeiro2012}   \\
    $0.60$       &  $0.390\pm0.063$             &  \cite{Blake2012}   \\
    $0.73$     &  $0.437\pm0.072$        &  \cite{Blake2012}   \\
    $0.77$     &  $0.490\pm0.018$        &  \cite{Song2009},\cite{LGuzzo2008]}   \\
    $0.80$     &  $0.470\pm0.080$        &  \cite{Torre2013}   \\
\hline
\end{tabular}
\caption{\label{tab:R} The growth data $R_{obs}(z)$.}
\end{table}

\subsection{Establishing the Equation of Consistency Relation}

In order to test general relativity on cosmological scales, T. Chiba and R. Takahashi\cite{Chiba2007} provided theoretically a consistency relation in general relativity which relates the luminosity distance and the density perturbation and relies on the matter density parameter today.  In this paper, we would like to provide a new way to test the consistency between the kinematic and dynamical probes in the framework of general relativity.

Solving Eq.~\ref{eq:defidensitydelta}, we can
obtain an analytical solution on $\delta(z)$ given by
\begin{eqnarray}\label{eq:deltasolution}
\delta(z)=\frac{5\Omega_{M0}E(z)}{2}\int^\infty_z\frac{1+z}{E^3(z)}dz,
\end{eqnarray}
where $E(z)=H(z)/H_0$. According to the definition of growth rate $f(z)$ and $f\sigma_8$ (Eq.~\ref{eq:fdefinition} and Eq.~\ref{eq:fsigma8definition}), we can derive the more specific forms:
\begin{eqnarray}\label{eq:Handfsigma8}
f\sigma_8(z)&=&\frac{5\Omega_{M0}H^2_0\sigma_8(0)}{2\delta(0)}(1+z)[\frac{1+z}{H^2(z)}
-\frac{dH(z)}{dz}\int^\infty_z\frac{1+z}{H^3(z)}dz],
\end{eqnarray}
\begin{eqnarray}\label{eq:Handf}
f(z)&=&-(1+z)[\frac{\frac{dH(z)}{dz}}{H(z)}+\frac{\frac{-(1+z)}{H^3(z)}}{\int^\infty_z\frac{1+z}{H^3(z)}dz}],
\end{eqnarray}
where we could see that there is no annoying parameter to be preset at the beginning for the expression of $f(z)$ and it can be used for predicting. As for our consistency testing, we changed the appearance of Eq.~\ref{eq:Handfsigma8} to make the terms whose value can be confirmed by simulation moved to one side of the equation:
\begin{eqnarray}\label{eq:Handfsigma8co1}
C(z)&=&\frac{f\sigma_8(z)}{H^2_0(1+z)[\frac{1+z}{H^2(z)}-\frac{dH(z)}{dz}\int^\infty_z\frac{1+z}{H^3(z)}dz]}
=\frac{5\Omega_{M0}\sigma_8(0)}{2\delta(0)}=\lambda,
\end{eqnarray}
We refer to it as the first consistency relation we get, using the kinematic and dynamical probes in the framework of general relativity. It is easy for comprehension that in this equation $R(z) (f\sigma_8(z))$ act as the $data_{obs}$ and $H^2_0(1+z)[\frac{1+z}{H^2(z)}-\frac{dH(z)}{dz}\int^\infty_z\frac{1+z}{H^3(z)}dz]$ is treated as $data_{cal}$. The ratio of them is the constant $\lambda$ but there is no need to preset it. The detailed processes will be illuminated in the next section. $C(z)$ have to be constant for all redshifts, as a null-test. $H^2_0$ has also been moved to left because it was a part of the simulated $H(z)$ data.

We also provided another type of consistency relation. The integral term can be set alone as:
\begin{eqnarray}\label{eq:Handfsigma8co2pro}
I(z)&=&\int^\infty_z\frac{1+z}{H^3(z)}dz
=\frac{1+z}{H'(z)H^2(z)}-\frac{f\sigma_8(z)}{\lambda H^2_0(1+z)H'(z)},
\end{eqnarray}
\begin{eqnarray}\label{eq:Handfsigma8co2pro2}
I(z_2)-I(z_1)&=&\int^{z_1}_{z_2}\frac{1+z}{H^3(z)}dz=[\frac{1+z}{H'(z)H^2(z)}-\frac{f\sigma_8(z)}{\lambda H^2_0(1+z)H'(z)}]\mid^{z_2}_{z_1},
\end{eqnarray}
\begin{eqnarray}\label{eq:Handfsigma8co2}
C(z_1;z_2)&=&\frac{\frac{f\sigma_8(z)}{H^2_0(1+z)H'(z)}\mid^{z_2}_{z_1}}{[\frac{1+z}{H'(z)H^2(z)}+\int\frac{1+z}{H^3(z)}dz]\mid^{z_2}_{z_1}}
=\frac{5\Omega_{M0}\sigma_8(0)}{2\delta(0)}=\lambda,
\end{eqnarray}
It is obvious that this equation enjoys a pleasing advantage that the calculated result can
only be dominated by data in observed range, which means that the errors introduced when tracing $H(z)$ of $z\rightarrow\infty$ can be avoided. However, it would suffer from the newborn errors coming from the increased calculation steps. The similar problem occurs when we change Eq.~\ref{eq:Handfsigma8co2pro2} to
\begin{eqnarray}\label{eq:Handfsigma8co3pro}
\lim\limits_{\Delta z\rightarrow0}{\frac{\int^{z_1}_{z_2}\frac{1+z}{H^3(z)}dz}{z_2-z_1}}&=&\lim\limits_{
\Delta z\rightarrow0}{\frac{[\frac{1+z}{H'(z)H^2(z)}-\frac{f\sigma_8(z)}{\lambda H^2_0(1+z)H'(z)}]\mid^{z_2}_{z_1}}{z_2-z_1}},
\end{eqnarray}
in which $\Delta z=z_2-z_1$ and $\lambda=\frac{5\Omega_{M0}\sigma_8(0)}{2\delta(0)}$,
\begin{eqnarray}\label{eq:Handfsigma8co3pro2}
-\frac{1+z}{H^3(z)}&=&\frac{d[\frac{1+z}{H'(z)H^2(z)}-\frac{f\sigma_8(z)}{\lambda H^2_0(1+z)H'(z)}]}{dz},
\end{eqnarray}
\begin{eqnarray}\label{eq:Handfsigma8co3}
C(z)&=&\frac{d[\frac{f\sigma_8(z)}{H^2_0(1+z)H'(z)}]/dz}{d[\frac{1+z}{H'(z)H^2(z)}]/dz+\frac{1+z}{H^3(z)}}=\lambda,
\end{eqnarray}
where the first derivative of $f\sigma_8(z)$ would be involved in to finish the calculation. It appears to be a noticeable problem when the quality of observed data still needs improvement. To be clear: errors produced by tracing $H(z)$ of $z\to\infty$ would not make test based on Eq.~\ref{eq:Handfsigma8co1} lose credibility, because $\frac{1+z}{H^3(z)}$ drops quickly as $z$ getting larger and the high-$z$ integral value is a negligible portion of the total result.

In this work, we put Eq.~\ref{eq:Handfsigma8co1} and Eq.~\ref{eq:Handfsigma8co2} into practice. We give them code names ``CRO (Consistency Relation using Original formula, for Eq.~\ref{eq:Handfsigma8co1})'' and ``CRI (Consistency Relation using Interval integral method, for Eq.~\ref{eq:Handfsigma8co2})'' for future reference. Furthermore, It is worthy discussing that how should we realize the CRI testing. Binning all data would smooth the fluctuation and reduce the sensitivity. So we choose to bin the last two points (because they are quite close to each other and represent the value of higher redshift) and set the result as the base point $(z_1,f\sigma_8(z_1))$ used for subtracting process.

We calculate $C(z)$(henceforward $C_{obs}(z))$ with
its theoretical uncertainty $\sigma_{C_{obs}}$ by means of $f\sigma_8$ and Hubble parameter $H(z)$ obtained from above mentioned two methods. The uncertainty $\sigma_{C_{obs}}$ is given as
\begin{eqnarray}\label{eq:deltaCtheo}
\sigma^2_{C_{obs}}&=&(\frac{\partial C}{\partial H})^2\sigma_H^2+(\frac{\partial C}{\partial H'})^2\sigma_{H'}^2+(\frac{\partial C}{\partial {f\sigma_8}})^2\sigma_{f\sigma_8}^2+(\frac{\partial C}{\partial {H_0}})^2\sigma_{H_0}^2\nonumber\\
&&+2(\frac{\partial C}{\partial H})(\frac{\partial C}{\partial {H'}})Cov(H,H')\nonumber\\
&&+2(\frac{\partial C}{\partial H})(\frac{\partial C}{\partial {f\sigma_8}})Cov(H,{f\sigma_8})\nonumber\\
&&+2(\frac{\partial C}{\partial H'})(\frac{\partial C}{\partial {f\sigma_8}})Cov(H',{f\sigma_8}).
\end{eqnarray}

Since we have already get the expressions of the consistency relation with the kinematic and dynamical probes, in the next section, we will discuss quantitatively how to use them to test general relativity on cosmological scales.
\vskip 0.5cm

\section{ Analysis methods, results and discussion }\label{section:analysis methods}

To quantify the relation between the kinematic and dynamical probes, we perform the method of parameterizations which has witnessed boom in recent years. Originally it is in the works about the state parameter of dark energy models\cite{Huterer2001,Efstathiou1999,Chevallier2001,Linder2003}. Inspired by their work, Holanda et al.\cite{Holanda2011} assumed $D_L(1+z)^{-2}/{D_A}=\eta(z)$, where $\eta(z)$ quantifies a possible epoch-dependent departure from the standard photon conserving scenario ($\eta=1$), and set two parametric presentations of $\eta(z)$ as $1+\eta_0z$ and $1+\eta_0z/(1+z)$ \cite{Chevallier2001} to test the DD relation. The former is a continuous and smooth linear expansion and the latter includes a possible epoch-dependent correction, making the value of $\eta$ become bounded when redshift goes higher. The parametric expressions enjoy many advantages. They have good sensitivity to observational data and avoid the divergence at extremely high $z$, which make them more useful when data in higher redshift become available. Since then, the method has been developed with more expressions and into two-dimension \cite{Zhengxiang 2011,Remya 2011,Meng 2012}. They helped to deal with a lot of significant work.

We have already seen that if general relativity is the correct theory of gravity in the universe, the equation
\begin{eqnarray}
C_{cal}(z)=\frac{5\Omega_{M0}\sigma_8(0)}{2\delta(0)}=constant,
\end{eqnarray}
should hold. It is also clear that the value of this constant can't be obtained based on the existing information in our work. With a view to the limitation from one-dimensional parameterization, two-dimensional parameterizations of $C(z)$, which is more general are enabled. For the possible redshift dependence of the consistency relation, we parameterize $C(z)$ as following:
\begin{subequations}\label{eta12}
\begin{align}
\label{eq:eta12 1}
C(z)&=C_1+C_2z ,\\
\label{eq:eta12 2}
C(z)&=C_1+C_2z/(1+z),\\
\label{eq:eta12 4}
C(z)&=C_1-C_2{\rm{ln}}(1+z),\\
\label{eq:eta12 5}
C(z)&=C_1-C_2{\rm{sin}}(z).
\end{align}
\end{subequations}
where the expression Eq.~\ref{eq:eta12 5} using trigonometric function is firstly proposed in the testing by us. As for all the expressions, we have $C=C_1$ when $z\ll1$. Moreover, the charming waving character of trigonometric function make it deserved to be tried in consistency parameterizations.

To determine the most probable values of the parameters in $C(z)$, maximum likelihood estimation is employed, via $L{\propto}e^{\chi^2}$ and
\begin{eqnarray}\label{eq:kai}
\chi^2=\sum_{z}\frac{[C(z)-C_{obs}(z)]^2}{\sigma^2_{C_{obs}}},
\end{eqnarray}
and the uncertainty $\sigma_{C_{obs}}$ is given by Eq.~\ref{eq:deltaCtheo}.

If the consistency relation holds using the kinematic and dynamical probes in the
framework of general relativity, the likelihood $e^{-\chi^2/2}$ would peak at $(C_1,C_2)=(\lambda,0)$, where $\lambda=\frac{5\Omega_{m0}\sigma_8(0)}{\delta(0)}$ for two-dimensional parameterizations. To draw the likelihood contours at 1, 2, 3$\sigma$ confidence level (CL), there is $\Delta\chi^2=$2.3, 6.17 and 11.8 for 1, 2 and 3$\sigma$ CL respectively. In order to obtain the value of $\Omega_{m0}\frac{\sigma_8(0)}{\delta(0)}$ simultaneously, we take the average of the $\lambda_{obs}$s (the value of $C_1$ where $e^{-\chi^2/2}$ peak at) as:
\begin{eqnarray}\label{eq:binlambda}
\bar{\lambda}_{obs}
&=&\frac{\sum_{i}(\lambda_{obsi}/\sigma^2_{\lambda_{obsi}})}{\sum_{i}1/\sigma^2_{\lambda_{obsi}}},\sigma^2_{\bar{\lambda_{obs}}}=\frac{1}{\sum_{i}1/\sigma^2_{\lambda_{obsi}}},
\end{eqnarray}
where $\lambda_{obsi}$ represents the $i\rm{th}$ result of the $\lambda_{obs}$, and $\sigma_{\lambda_{obsi}}$ denotes its observational uncertainty.

\begin{table}[tbp]
\centering
\begin{tabular}{|lcc|}
\hline
{$parameterization$}   & $\chi^2/d.o.f.$ & $C_1 \& C_2$   \\
\hline
                   $$    & Method A & $$  \\
\hline
$C=C_1+C_2z$  &  $0.1818$ &  $0.6226\pm 0.0830$   \\
$ $           & $ $         & $0.0232\pm 0.1655$\\
$C=C_1+C_2z/(1+z)$ &  $0.1830$        & $0.6320\pm 0.1046$ \\
$ $             & $ $                   & $0.0044\pm 0.3285$\\
$C=C_1-C_2{\rm{ln}}(1+z)$     &  $0.1826$  &  $0.6267\pm 0.0933$ \\
$ $                           & $ $          & $-0.0181\pm 0.2362$\\
$C=C_1-C_2{\rm{sin}}(z)$  &  $0.1824$  &  $0.6256\pm 0.0879$ \\
$ $                      & $ $           & $-0.0177\pm 0.1870$\\
$\Omega_{M0}\frac{\sigma_8(0)}{\delta(0)}=0.2533\pm 0.0540$ & $$ & $$  $$\\
\hline
        $$ & Method B&   $$   \\
\hline
$C=C_1+C_2z$  &  $0.1505$ &  $0.5054\pm 0.1000$   \\
$ $           & $ $          & $0.2930\pm 0.2071$\\
$C=C_1+C_2z/(1+z)$ &  $0.1786$        & $0.4839\pm 0.1281$ \\
$ $                & $ $                & $0.5123\pm 0.4112$\\
$C=C_1-C_2{\rm{ln}}(1+z)$     &  $0.1641$  &  $0.4937\pm 0.1133$ \\
$ $                           & $ $          & $-0.3946\pm 0.2954$\\
$C=C_1-C_2{\rm{sin}}(z)$  &  $0.1597$  &  $0.5012\pm 0.1062$ \\
$ $                       & $ $          & $-0.3179\pm 0.2335$\\
$\Omega_{M0}\frac{\sigma_8(0)}{\delta(0)}=0.2546\pm 0.0625$ & $$ & $$  $$\\
\hline
\end{tabular}
\caption{\label{table:parameterizationCRO} Marginal mean and standard deviation of the results of CRO (Eq.~\ref{eq:Handfsigma8co1}) as inferred from $\chi^2/d.o.f.$ by method A and B respectively.}
\end{table}

\begin{table}[tbp]
\centering
\begin{tabular}{|lcc|}
\hline
{$parameterization$}   & $\chi^2/d.o.f.$ & $C_1 \& C_2$   \\
\hline
                   $$    & Method A & $$  \\
\hline
$C=C_1+C_2z$  &  $0.1226$ &  $0.7098\pm 0.2048$   \\
$ $           & $ $         & $-0.6219\pm 0.6169$\\
$C=C_1+C_2z/(1+z)$ &  $0.1303$        & $0.7233\pm 0.2287$ \\
$ $             & $ $                   & $-0.9098\pm 0.9631$\\
$C=C_1-C_2{\rm{ln}}(1+z)$     &  $0.1263$  &  $0.7174\pm 0.2168$ \\
$ $ & $ $                                    & $0.7602\pm 0.7774$\\
$C=C_1-C_2{\rm{sin}}(z)$  &  $0.1242$  &  $0.7117\pm 0.2086$ \\
$ $ & $ $                                & $0.6448\pm 0.6479$\\
$\Omega_{M0}\frac{\sigma_8(0)}{\delta(0)}=0.2105\pm 0.1391$ & $$ & $$  $$\\
\hline
                  $$    & Method B &  $$     \\
\hline
$C=C_1+C_2z$  &  $0.0204$ &  $0.5808\pm 0.3419$   \\
$ $ & $ $                   & $-0.4406\pm 1.0140$\\
$C=C_1+C_2z/(1+z)$ &  $0.0217$        & $0.5907\pm 0.3801$ \\
$ $ & $ $                               & $-0.6496\pm 1.5967$\\
$C=C_1-C_2{\rm{ln}}(1+z)$     &  $0.0211$  &  $0.5864\pm 0.3612$ \\
$ $ & $ $                                    & $0.5407\pm 1.2794$\\
$C=C_1-C_2{\rm{sin}}(z)$  &  $0.0207$  &  $0.5823\pm 0.3481$ \\
$ $ & $ $                               & $0.4578\pm 1.0659$\\
$\Omega_{M0}\frac{\sigma_8(0)}{\delta(0)}=0.1799\pm 0.2417$ & $$ & $$  $$\\
\hline
\end{tabular}
\caption{\label{table:parameterizationCRI} Marginal mean and standard deviation of the results of CRI (Eq.~\ref{eq:Handfsigma8co2}) as inferred from $\chi^2/d.o.f.$ by method A and B respectively.}
\end{table}

\begin{figure*}[t,m,b]
\begin{center}
\begin{tabular}{cc}
\includegraphics*[width=75.3mm,height=6cm]{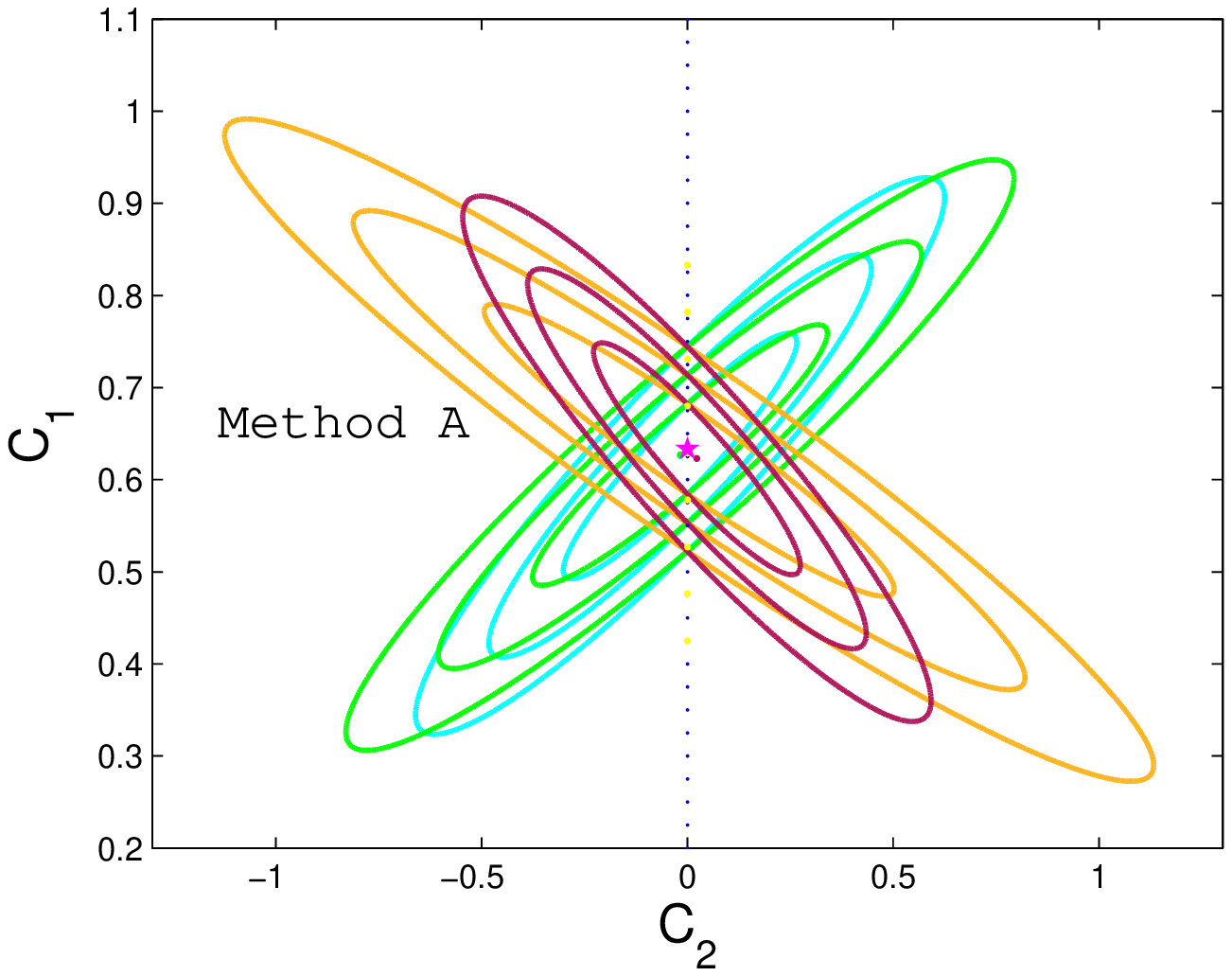}&
\includegraphics*[width=75.3mm,height=6cm]{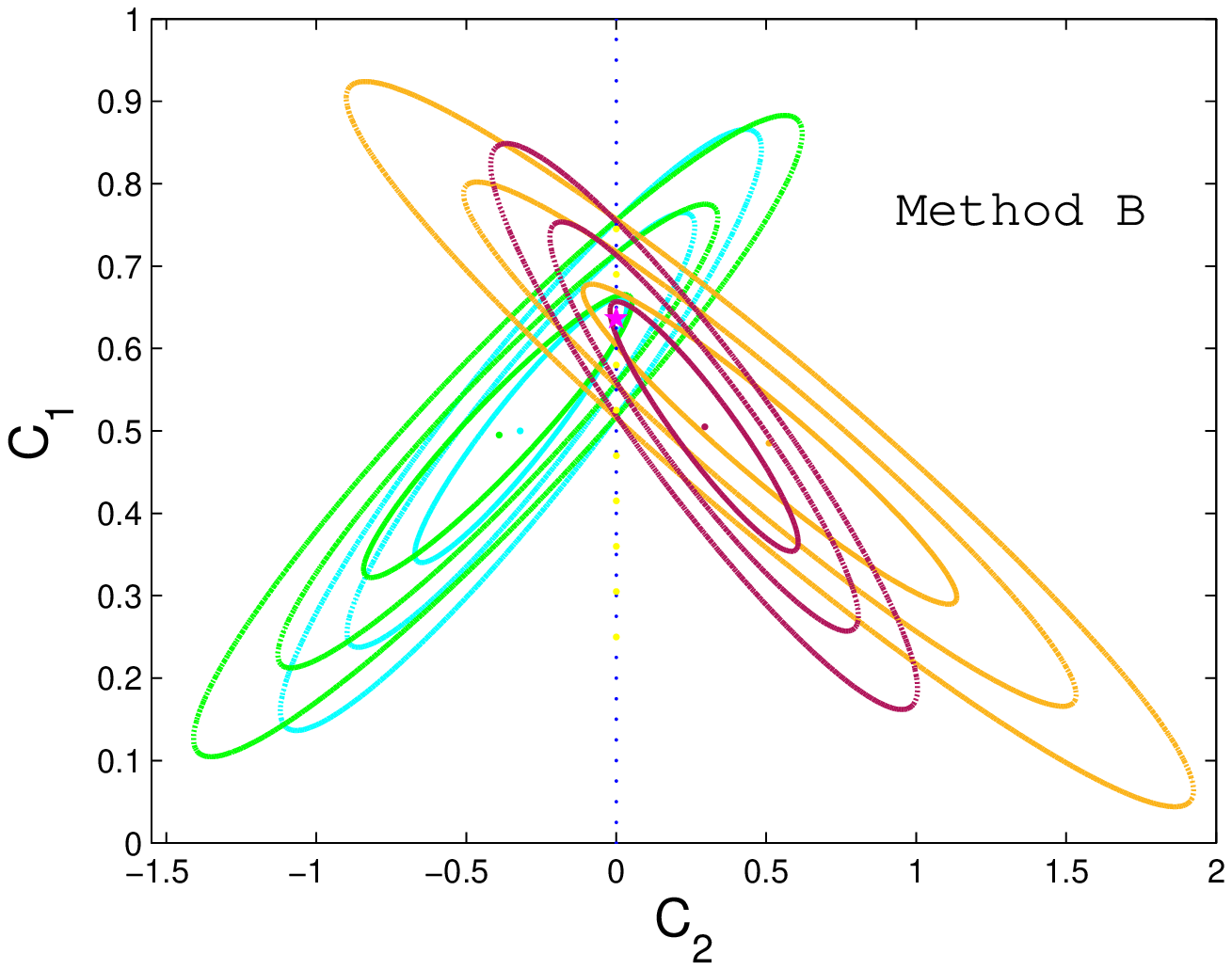}
\end{tabular}
\caption{\label{fig:2Dorigin}Marginal contours in the $C_2-C_1$ plane of two-dimensional parameterizations for CRO (Eq.~\ref{eq:Handfsigma8co1}). The contours lay inside-out denote the boundaries of $1\sigma$, $2\sigma$ and $3\sigma$ CL,  respectively. Four color lines denote four two-dimensional parameterizations. Meanwhile the dots located in the centre of the contours in corresponding colors denote the best fits for parameterizations. Moreover, the vertical dashed line stands for $C_2=0$ and the pink pentagram stands for the value ($\bar{\lambda}_{obs}$,0) of ($C_1, C_2$) for reference.}
\end{center}
\end{figure*}

\begin{figure*}[t,m,b]
\begin{center}
\begin{tabular}{cc}
\includegraphics*[width=75.3mm,height=6cm]{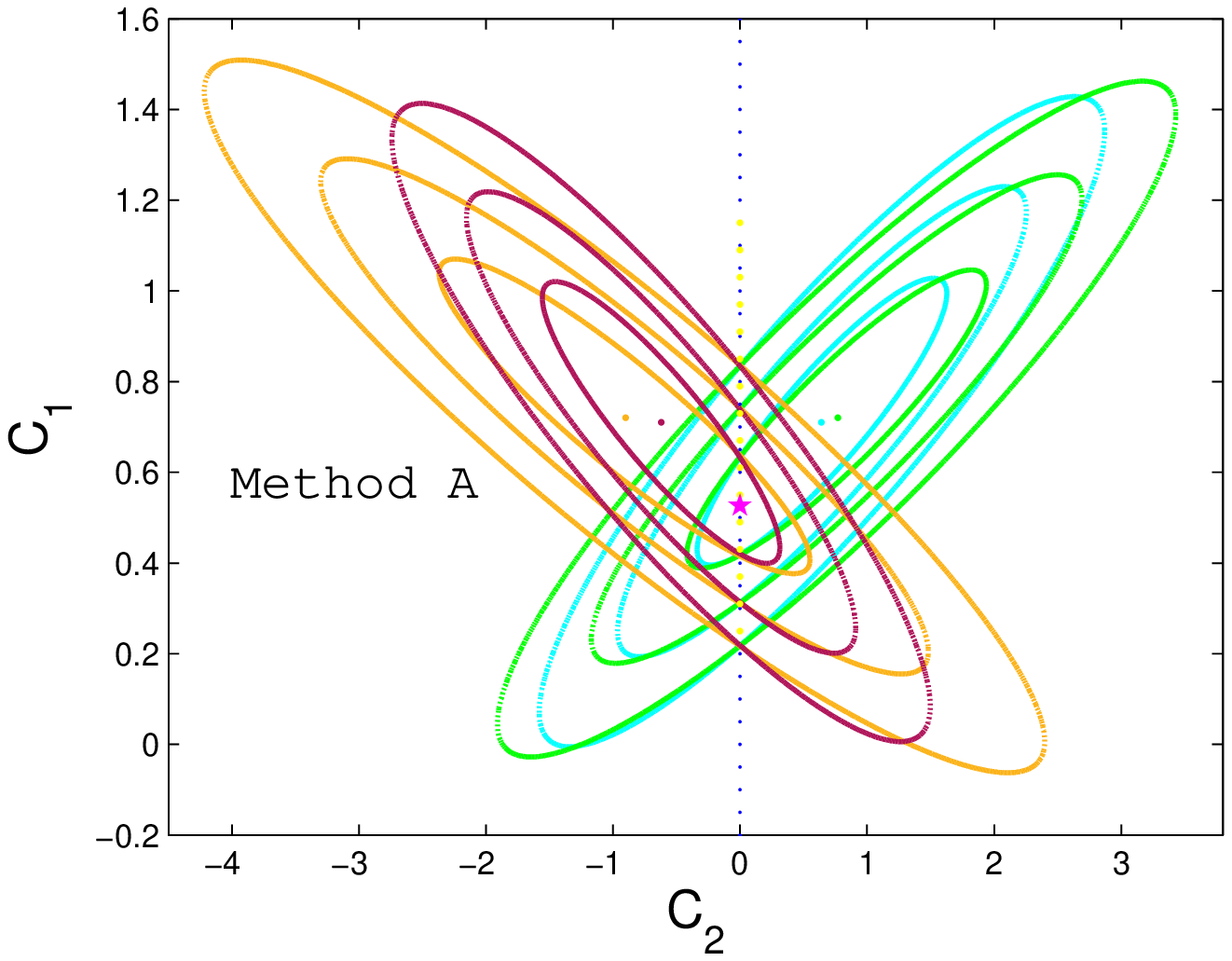}&
\includegraphics*[width=75.3mm,height=6cm]{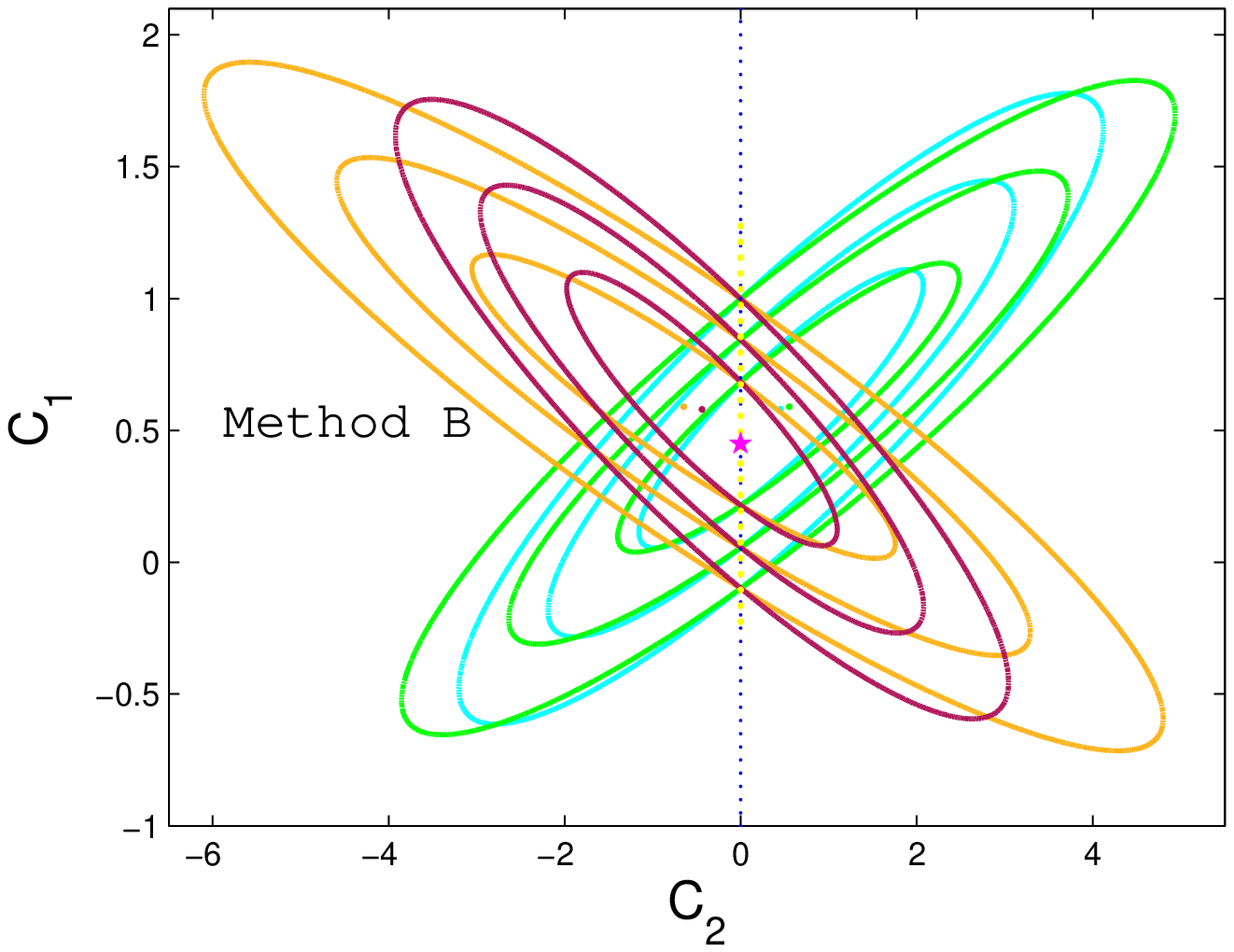}
\end{tabular}
\caption{\label{fig:2Dinterval}Marginal contours in the $C_2-C_1$ plane of two-dimensional parameterizations for CRI (Eq.~\ref{eq:Handfsigma8co2}). While $z_1$ has been fixed, horizontal axis shows $z_2$. The other comments are the same as Fig.~3.  }
\end{center}
\end{figure*}

Using the maximum likelihood described above, we have constrained the parameter values in parameterizations of $C(z)$. As for CRO and CRI, using the parametric method (Method A) and Gaussian Process (Method B) respectively,
the $\chi^2$ per degrees of freedom ($\chi^2 /d.o.f.$) \cite{Press 1994} and the best-fit value of $(C_1,C_2)$ with their errors are summarized in Table \ref{table:parameterizationCRO},\ref{table:parameterizationCRI}. The normalized likelihood distribution functions are plotted in Figure 3,4, where the different color curves correspond to the results from four two-dimensional parameterizations respectively, with the principle of correspondence as

$C(z)=C_1+C_2z\rightarrow$ red,

$C(z)=C_1+C_2z/(1+z)\rightarrow$ orange,

$C(z)=C_1-C_2\rm{ln}(1+z)\rightarrow$ green,

$C(z)=C_1-C_2\rm{sin}(z)\rightarrow$ blue.

According to the results shown in Figure 3,4 and Table \ref{table:parameterizationCRO},\ref{table:parameterizationCRI}, the consistency is clear between the kinematic and dynamical probes for all parameterizations within $1\sigma$ CL. CRI shows better result than CRO does, but it can't be ignored that its error gets larger due to the extra calculating steps. It means that in terms of the present condition of data, CRO implements the stricter constraint, and the calculation using our mock data for high-$z$ Hubble data doesn't deviate much from the real one. In conclusion, all two-dimensional parameterizations give a substantial support to the consistency relation between the kinematic and dynamical probes within $1\sigma$ CL.


\begin{figure*}[t,m,b]\label{fig:cocoplot}

\begin{center}
\begin{tabular}{cc}

\includegraphics*[width=75.3mm,height=6cm]{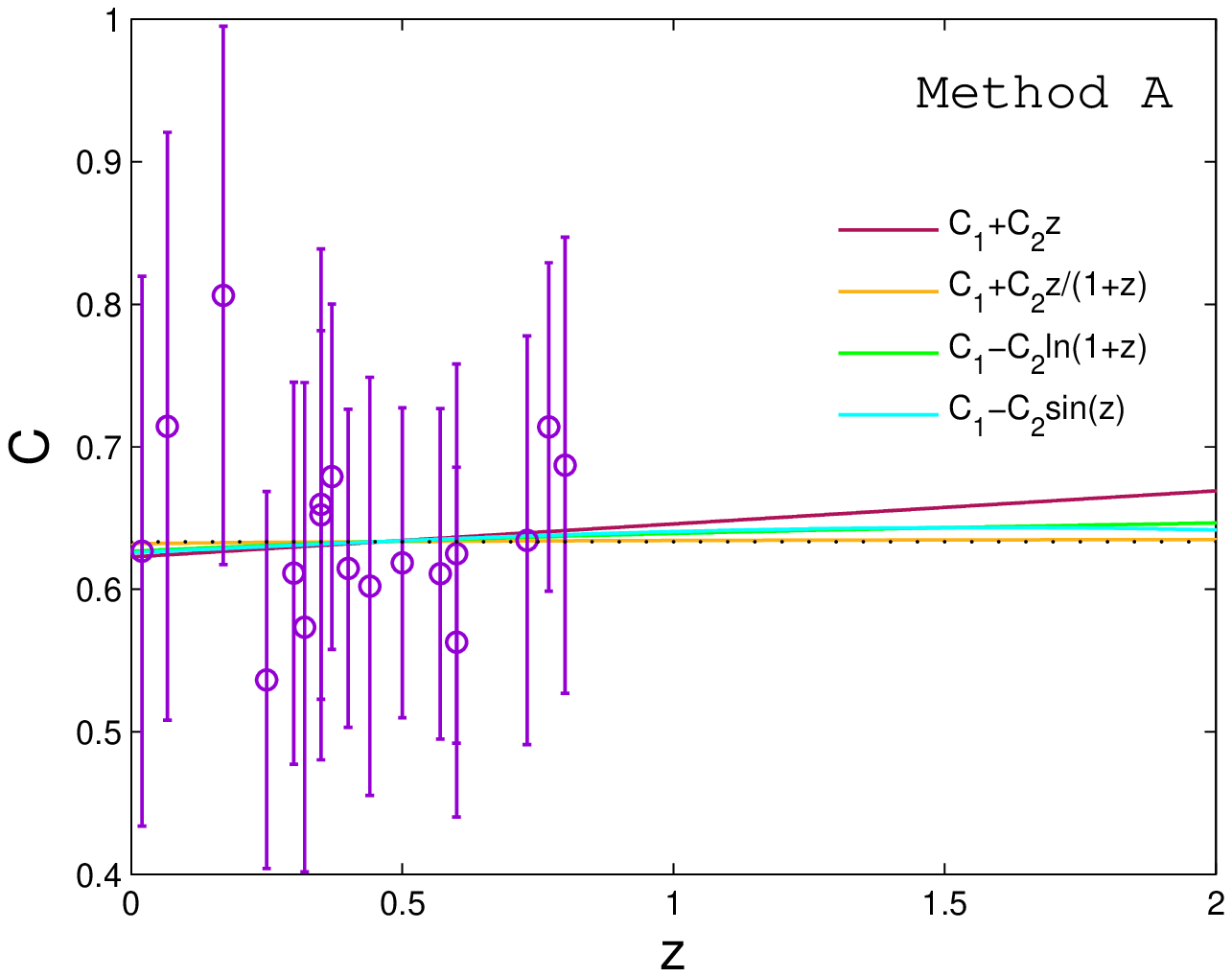}&
\includegraphics*[width=75.3mm,height=6cm]{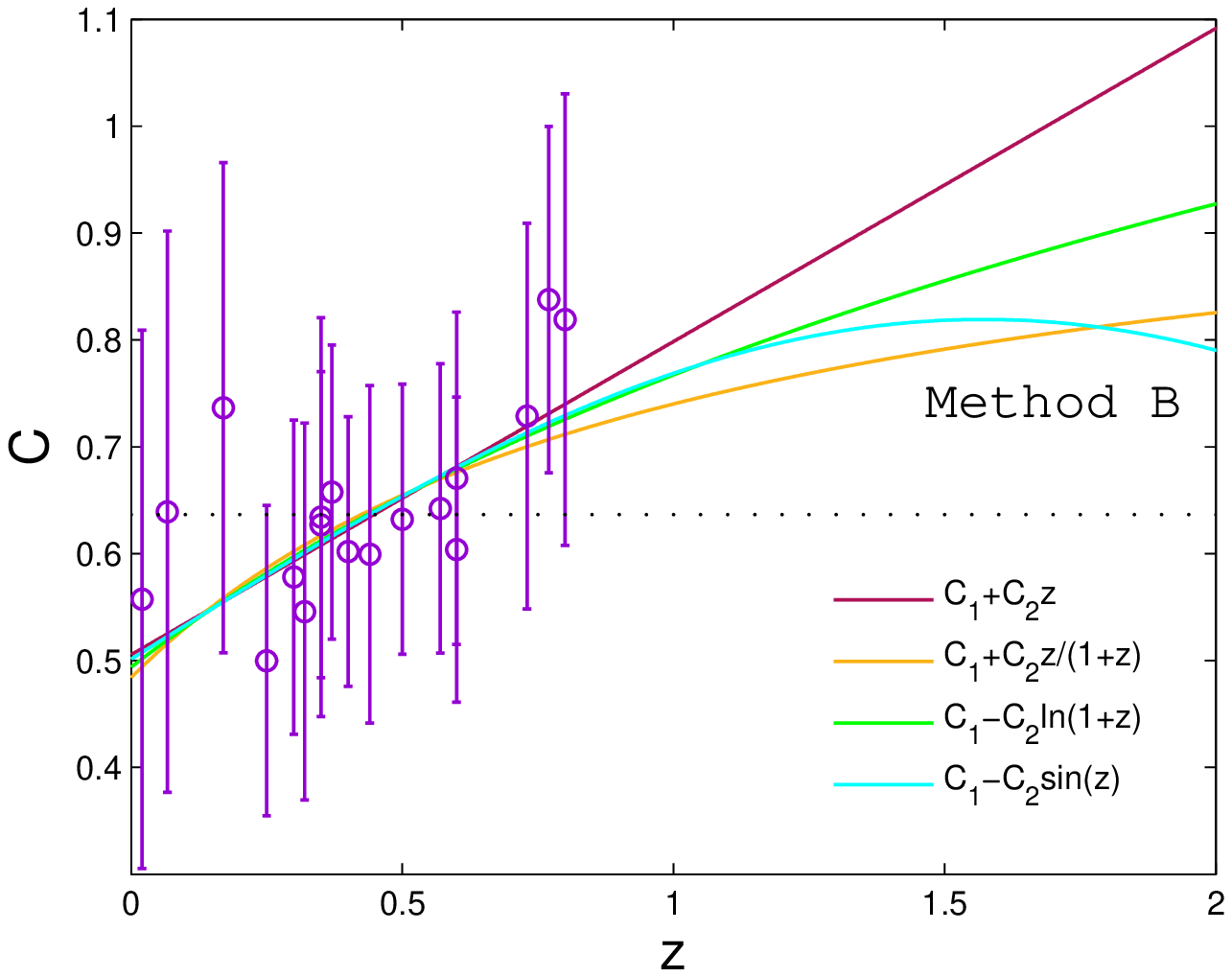}

\end{tabular}
\caption{Visualization of the results of CRO (Eq.~\ref{eq:Handfsigma8co1}). Displayed are two dimensional parameterizations using best fit values from Table III. The points and error bars are calculated by using method A and B for the $R(z)$ dataset. The horizontal dashed lines show where $\bar{\lambda}_{obs}$(Eq.~\ref{eq:binlambda}) locates.  }

\end{center}
\end{figure*}

In order to understand behaviour of the used parameterizations in a quantitative manner, we plot the result of CRO, which does stricter constraint, using its best fit values along with $C_{obs}(z)$ and $\sigma_{C_{obs}}$  in Figure 5 from the two methods: Method A and B. For the parametric method (Method A), fitting curves are pleasant thanks to good parametrization of $H(z)$, so there is not obvious difference. By means of the non-parametric method (Method B) from Gaussian Processes (GP), it is clear from the Figures 5 that parametrization forms $C(z)\sim z/(1+z)$ (Eq.~\ref{eq:eta12 2}) and $C(z)\sim {\rm{sin}}(z)$ (Eq.~\ref{eq:eta12 5}) are good choices for modeling the consistency relation because other two  parameterizations gradually deviate from $\bar{\lambda}_{obs}$ (Eq.~\ref{eq:binlambda}) as the increase of redshift. As for $C(z)=C_1+C_2z$ (Eq.~\ref{eq:eta12 1}), there is a large slope, and no chance for it to be convergent. $C(z)=C_1-C_2{\rm{ln}}(1+z)$ (Eq.~\ref{eq:eta12 4}) inclines too much as well. The four of them give approached departures at $z=0$ (which for Eq.~\ref{eq:eta12 1} and Eq.~\ref{eq:eta12 4} are slightly smaller). In sum, as seen in Figure 5, $C(z)\sim z/(1+z)$ and $C(z)\sim {\rm{sin}}(z)$ perform better than the other two.

What is noteworthy is that the parametrization model using trigonometric function is considered aesthetically very beautiful. It provides an error model which is always dipping and heaving around the expected value. We can see that it performs well in this work, providing a tight constraint to the expected value. Therefore, in consideration of the pivotal role of growth data in high redshift region, we suggest to employ the parametrization forms $C(z)\sim z/(1+z)$ and $C(z)\sim {\rm{sin}}z$ in future testing of consistency relations between
the kinematic and dynamical probes in the framework of general relativity.

\vskip 0.5cm
\section{Validity check for the testing method with mock data}

It is worth stressing that our method testing general relativity on cosmological scale is independent of any cosmological model. But $f\sigma_8$, the dynamical probe data we used, which can be considered as "almost model-independent", are also not evaluated in completely model-independent processes. That is to say, the testing may prefer the models which participated in the method of providing data (for $f\sigma_8$ it is $\Lambda$CDM model)\cite{Nesseris 2015}. It occurred to us that, in order to confirm that the validity of this test wasn't whelmed by model-dependency, we should have it checked.

The main weak point of cosmological models based on the validity of general relativity lies in involving fined tuned parameters, such as standard cold dark matter (SCDM)\cite{Nesseris 2004}, cold dark matter with cosmological constant $\Lambda$ ($\Lambda$CDM, also called LCDM model)\cite{Peebles 2003}, quiessence model (QUIES)\cite{Alam 2003}, dark energy with equation of state $w(z) = w_0 + w_1z$ (Linear)\cite{Huterer2001}, $w(z) = w_0 + w_1z/(1 + z)$ (CPL)\cite{Chevallier2001}. Since all of them are based on general relativity, they should enjoy significant advantage compared to other models which disfavor general relativity, like DGP (Dvali-Gabadadze-Poratti) model. Table \ref{table:modelcomparing} presents the best-fit parameters of above-mentioned six cosmological models using currently available OHD dataset. The comparison of the observed and theoretical evolution of the Hubble parameter $H(z)$ is given in the top-left panel in Fig.~7. Because of the significantly larger $\chi^2_{min}/d.o.f$, we conclude that SCDM does not provide a good fit to $H(z)$ data. By means of the best-fit parameters of cosmological models, Eq.~\ref{eq:Handf} and Eq.~\ref{eq:Handfsigma8co1}, the theoretical results of the growth rate $f(z)$ and $f\sigma_8/\lambda$ are visualized in Fig.~6. It is indicated that four cosmological models except of SCDM are visually consistent with the reasonable trend.

It can be deduced that, if the validity was whelmed by interference from the model ($\Lambda$CDM, which $f\sigma_8$ depends on), the success of the test would depend on the similarity between the $H(z)$ data we used and those predicted by $\Lambda$CDM model. Potential inconsistencies between the predicted growth data and the observed ones are introduced, which are not due to failure of what we test but the inconsistent use of $H(z)$. In order to check the validity, we make DGP model involved in. Moreover, in Fig.~7 we compared its fitting result of $H(z)$ with the parametrization of $H(z)$ of method A (Eq.~\ref{eq:Hz}), and four models which we found more reasonable in the earlier paragraph. We can see that fitting result of $H(z)$ in DGP model is quite approached to the $\Lambda$CDM case. However, since DGP is base on a different principle from general relativity, the consistency relations they use wouldn't be the same. Furthermore, according to predictions of future observation\cite{Weinberg2013}, the relative uncertainty of these data can be reduced up to 1\% to set stricter constraints. Taking these into consideration, we test $H(z)$ fitting result in DGP model with principles of general relativity and DGP model respectively, to see how much difference there would be.

DGP (Dvali-Gabadadze-Poratti) model is a model of brane world where three-dimensional brane is embedded in an infinite five-dimensional spacetime (bulk). The action for this five-dimensional theory is:
\begin{eqnarray}\label{eq:DGPS}
S&=&\frac{1}{2}M^3_{(5)}\int d^4xdy\sqrt{-g_{(5)}}R_{(5)}\nonumber\\
&&\frac{1}{2}M^2_{(4)}\int d^4x\sqrt{-g_{(4)}}R_{(4)}+S_m,
\end{eqnarray}
in which subscripts 4 represents the quantities on the brane and subscripts 5 denotes those in the bulk. $M_{(4)}$ and $M_{(5)}$ denote four and five-dimensional reduced Planck mass respectively. $S_m$ represents the action of the matter on the brane.

The consistency relations based on general relativity hold on cosmological scales when dark energy has anisotropic pressure or interaction with dark matter\cite{Kunz 2007}. The modification of gravity theories affects the gravitational instability. As for Eq.~\ref{eq:defidensitydelta}, it gets the third term, the self-gravity of density perturbations modified. In scalar-tensor theories of gravity, Eq.~\ref{eq:defidensitydelta} is modified as:
\begin{eqnarray}\label{eq:defidensitydeltaDGP}
\ddot{ \delta}+2H\dot{\delta}-4{\pi}G_{eff}\rho_M{\delta} &=& 0,
\end{eqnarray}
in which $G_{eff}$ denotes the effective local gravitational ``constant'', which is time dependent, measured by Cavendish-type experiment. In general it may be written as:
\begin{eqnarray}\label{eq:defidensitydeltaDGP2}
\ddot{ \delta}+2H\dot{\delta}-4{\pi}G\rho_M(1+\frac{1}{3\beta}){\delta} &=& 0,
\end{eqnarray}
in which $\beta$ in general depends on time. Once we specify the modified gravity theory, it is determined. The evolution of density perturbations during time is also modified. Furthermore, the Friedmann equation is changed to:
\begin{eqnarray}\label{eq:FreDGP1}
H^2+\frac{K}{a^2}=\frac{8\pi G}{3}(\sqrt{\rho_m+\rho_{r_c}}+\sqrt{\rho_{r_c}})^2,
\end{eqnarray}
in which $\rho_{r_c}=3/(32\pi Gr^2_c)$ and $r_c$ can be given as $r_c=M^2_{(4)}/2M^3_{(5)}$.
$\beta$ of Eq.~\ref{eq:defidensitydeltaDGP2} can be obtained by:
\begin{eqnarray}\label{eq:DGPbeta}
\beta=1-2r_cH(1+\frac{\dot{H}}{3H^2}),
\end{eqnarray}
If $\Omega_{r_c}=\frac{1}{4r^2_cH^2_0}$ involved in, Eq.~\ref{eq:FreDGP1} changes to\cite{Wan 2007}:
\begin{eqnarray}\label{eq:DGPHz}
H^2(z)/H^2_0&=&\Omega_{k0}(1+z)^2+[\sqrt{\Omega_{r_c}}\nonumber\\
&&+\sqrt{\Omega_{r_c}+\Omega_{m0}(1+z)^3}]^2,
\end{eqnarray}
and
\begin{eqnarray}\label{eq:DGPHz1}
\Omega_{k0}+[\sqrt{\Omega_{r_c}}+\sqrt{\Omega_{r_c}+\Omega_{m0}}]^2=1,
\end{eqnarray}
which can be obtained when $z=0$. Eq.~\ref{eq:DGPHz} and Eq.~\ref{eq:DGPHz1} can be used to get fitting result with observational Hubble data.

We choose CRO (Eq.~\ref{eq:Handfsigma8co1}) to be ``modified'' with the ``correction factor'': $(1+\frac{1}{3\beta})$, since it directly set comparison between $f\sigma_{8obs}$ and the calculated value of $f\sigma_{8}/\lambda$. As for solving Eq.~\ref{eq:defidensitydeltaDGP2}, we use iterative computing method by the fourth-order Runge-Kutta scheme. Since we are interested in the influence from the change of testing principle, the computing process shares the same initial conditions of matter density perturbations with $\lambda$CDM model\cite{Hirano 2015}. The comparison between the results of $\Lambda$CDM, ``DGP in GR''(which uses $H(z)$ fitting result in DGP and testing principle of general relativity) and DGP model are illustrated with Fig.~7 and detailed analysis information is listed in Table~\ref{table:parameterizationABC}. What we can see is that $\Lambda$CDM and ``DGP in GR'' pass the test with obvious advantage while DGP model gets apparent deterioration in the result compared with ``DGP in GR''. In other words, based on the same expression of $H(z)$ which provides quite similar data with $\Lambda$CDM model, the change of testing principle leads to significantly different results. Furthermore, because of the consistency between the $H(z)$ fitting result of $\Lambda$CDM and DGP model, the failure of DGP model to pass the test can't be imputed to the inconsistent use of $H(z)$. In conclusion, our testing principle, the consistency relation is the one which dominates the result.
\begin{table}[tbp]
\centering
\begin{tabular}{|lcccc|}
\hline
{$Model$}   & $H(z)$ & $\chi^2_{min}$ & $\chi^2_{min}/d.o.f.$ & Best fit parameters    \\
\hline
$      $   &  $    $    &  $    $    &  $          $  & $  $\\
$Linear$   &  $H^2(z)=H^2_0[\Omega_{0m}(1+z)^3+(1-\Omega_{0m})$      &  $18.3781 $ &  $0.5405 $ & $H_0=71.032\pm4.4611,$\\
$      $   &  $       \times(1+z)^{3(1+w_0-w_1)}e^{3w_1z}]$        &  $          $ &  $          $ & $\Omega_{0m}=0.2223\pm0.0210,$\\
$      $   &  $    $   &  $    $     &  $          $  & $w_0=-1.0081\pm0.2688,$\\
$      $   &  $    $   &  $    $     &  $          $  & $w_1=0.2454\pm0.1196$\\
$      $   &  $    $   &  $    $     &  $          $  & $  $\\
$CPL$   &  $H^2(z)=H^2_0[\Omega_{0m}(1+z)^3+(1-\Omega_{0m})$      &  $18.0807 $ &  $0.5318 $ & $H_0=72.714\pm5.6131,$\\
$      $   &  $       \times(1+z)^{3(1+w_0+w_1)}e^{3w_1(1/(z+1)-1)}]$  &  $          $ &  $          $  & $\Omega_{0m}=0.1870\pm0.1390,$\\
$      $   &  $    $    &  $    $    &  $          $  & $w_0=-1.1378\pm0.3693,$\\
$      $   &  $    $    &  $    $    &  $          $  & $w_1=1.0845\pm0.9859$\\
$      $   &  $    $    &  $    $    &  $          $  & $  $\\
$QUIES$  &  $H^2(z)=H^2_0[\Omega_{0m}(1+z)^3+(1-\Omega_{0m})$  & $18.4592$  & $0.5274$  &  $H_0=70.540\pm4.5097,$ \\
$     $  &  $\times(1+z)^{3(1+w)}]$  & $        $  & $        $  &  $\Omega_{0m}=0.2529\pm0.0261,$ \\
$      $   &  $    $    &  $    $    &  $          $  & $w=-1.0013\pm0.2388$\\
$      $   &  $    $    &  $    $    &  $          $  & $  $\\
$\Lambda CDM$     &  $H^2(z)=H^2_0[\Omega_{0m}(1+z)^3+(1-\Omega_{0m})]$    & $18.4592$    & $0.5128$ & $H_0=70.519\pm1.8259,$\\
$      $   &  $    $    &  $    $    &  $          $  & $\Omega_{0m}=0.2529\pm0.0251$\\
$      $   &  $    $    &  $    $    &  $          $  & $  $\\
$SCDM$  &  $H^2(z)=H^2_0(1+z)^3$   &  $175.0028$ &  $4.7298$ & $H_0=45.701\pm0.6292$\\
$      $   &  $    $    &  $    $    &  $          $  & $  $\\
$DGP$   &  $H^2(z)=H^2_0[\Omega_{k0}(1+z)^2+(\sqrt{\Omega_{r_c}}$      &  $18.4486 $ &  $0.5271 $ & $H_0=70.053\pm1.6963,$\\
$      $   &  $  +\sqrt{\Omega_{r_c}+\Omega_{m0}(1+z)^3}]$        &  $          $ &  $          $ & $\Omega_{m0}=0.2688\pm0.0208,$\\
$      $   &  $  ps: \Omega_{k0}+(\sqrt{\Omega_{r_c}}+\sqrt{\Omega_{r_c}+\Omega_{m0}})^2=1$        &  $          $ &  $          $ & $\Omega_{r_c}=0.1970\pm0.0151$\\
\hline
\end{tabular}
\caption{\label{table:modelcomparing} Marginal mean and standard deviation of model parameters in $H(z)$ expressions for various models as inferred from $\chi^2/d.o.f.$.}
\end{table}

\begin{figure*}[t,m,b]
\begin{center}
\begin{tabular}{cc}
\includegraphics*[width=75.3mm,height=6cm]{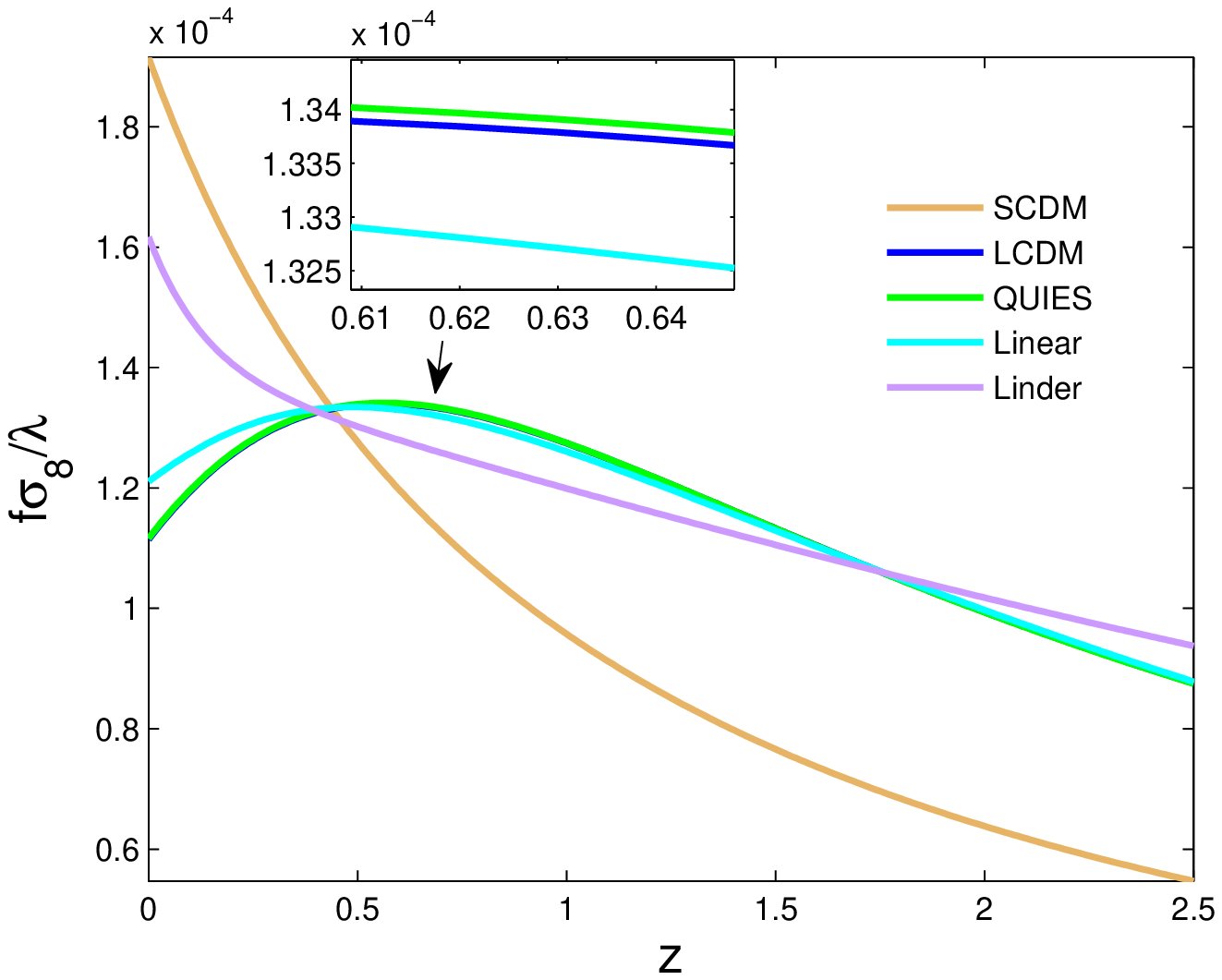}&
\includegraphics*[width=75.3mm,height=6cm]{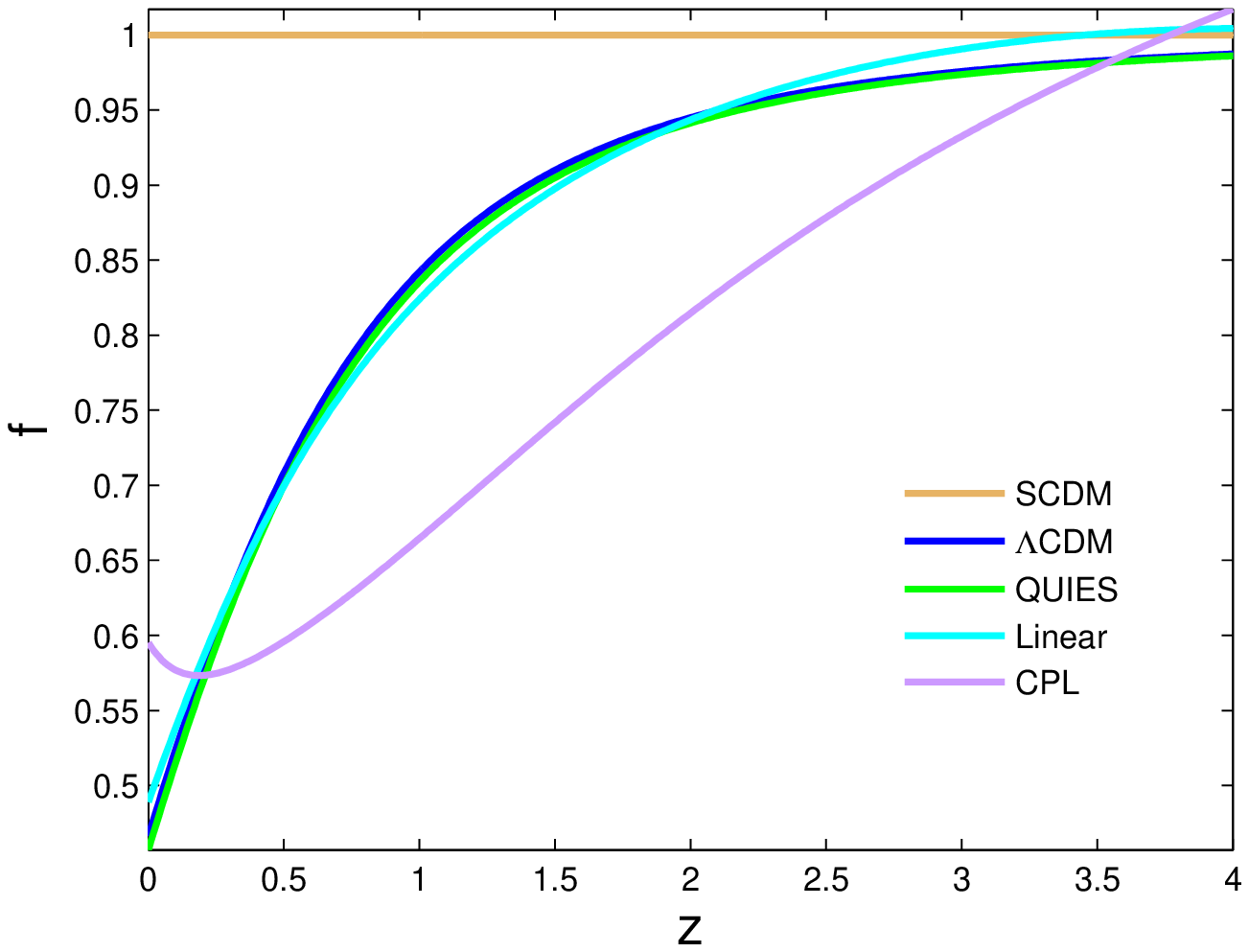}
\end{tabular}
\caption{\label{fig:simul}Prediction of the evolution of growth rate $f\sigma_8/\lambda$ (left panel) and $f(z)$ (right panel). Five color lines correspond to five cosmological models.}
\end{center}
\end{figure*}
\begin{table}[tbp]
\centering
\begin{tabular}{|lcc|}
\hline
{$parameterization$}   & $\chi^2/d.o.f.$ & $C_1 \& C_2$  \\
\hline
                   $$    & $\Lambda$CDM & $$  \\
\hline
$C=C_1+C_2z$  &  $0.3671$ &  $0.6320\pm     0.0564$   \\
$ $ & $ $                   & $0.0818\pm 0.1037$\\
$C=C_1+C_2z/(1+z)$ &  $0.3852$        & $0.6360\pm 0.0674$ \\
$ $ & $ $                               & $0.1164\pm 0.2020$\\
$C=C_1-C_2{\rm{ln}}(1+z)$     &  $0.3768$  &  $0.6334\pm 0.0618$ \\
$ $ & $ $                                    & $-0.1003\pm 0.1468$\\
$C=C_1-C_2{\rm{sin}}(z)$  &  $0.3741$  &  $0.6337\pm 0.0591$ \\
$ $ & $ $                                    & $-0.0836\pm 0.1171$\\
$\Omega_{M0}\frac{\sigma_8(0)}{\delta(0)}=0.2689\pm 0.0389$ & $$ & $$ $$\\
$\frac{\sigma_8(0)}{\delta(0)}=1.0632\pm 0.1742 $ & $$ & $$  $$\\
\hline
                   $$    & DGP in GR & $$  \\
\hline
$C=C_1+C_2z$  &  $0.3417$ &  $0.6587\pm     0.0580$   \\
$ $ & $ $                   & $0.0905\pm 0.1064$\\
$C=C_1+C_2z/(1+z)$ &  $0.3587$        & $0.6603\pm 0.0686$ \\
$ $ & $ $                               & $0.1384\pm 0.2058$\\
$C=C_1-C_2{\rm{ln}}(1+z)$     &  $0.3506$  &  $0.6591\pm 0.0632$ \\
$ $ & $ $                                    & $-0.1144\pm 0.1501$\\
$C=C_1-C_2{\rm{sin}}(z)$  &  $0.3483$  &  $0.6599\pm 0.0606$ \\
$ $ & $ $                                & $-0.0943\pm 0.1200$\\
$\Omega_{M0}\frac{\sigma_8(0)}{\delta(0)}=0.2813\pm 0.0407$ & $$ & $$  $$\\
$\frac{\sigma_8(0)}{\delta(0)}=1.0464\pm 0.1717$ & $$ & $$  $$\\
\hline
                  $$    & DGP &  $$     \\
\hline
$C=C_1+C_2z$  &  $0.3417$ &  $0.6588\pm 0.5799$   \\
$ $ & $ $                   & $0.0905\pm 0.1064$\\
$C=C_1+C_2z/(1+z)$ &  $0.3587$        & $0.6603\pm 0.0686$ \\
$ $ & $ $                               & $0.1384\pm 0.2058$\\
$C=C_1-C_2{\rm{ln}}(1+z)$     &  $0.3506$  &  $0.6591\pm 0.0632$ \\
$ $ & $ $                                    & $-0.1144\pm 0.1501$\\
$C=C_1-C_2{\rm{sin}}(z)$  &  $0.3483$  &  $0.6599\pm 0.0606$ \\
$ $ & $ $                               & $-0.0943\pm 0.1200$\\
$\Omega_{M0}\frac{\sigma_8(0)}{\delta(0)}=0.3369\pm 0.0496$ & $$ & $$ $$\\
$\frac{\sigma_8(0)}{\delta(0)}=1.2532\pm 0.2085$ & $$ & $$  $$\\
\hline
\end{tabular}
\caption{\label{table:parameterizationABC} Marginal mean and standard deviation of results for $\lambda$CDM, ``DGP in GR'' and DGP model respectively as inferred from $\chi^2/d.o.f$.}
\end{table}

\begin{figure*}[t,m,b]\label{fig:mocking}
\begin{center}
\begin{tabular}{cc}
\includegraphics*[width=75.3mm,height=6cm]{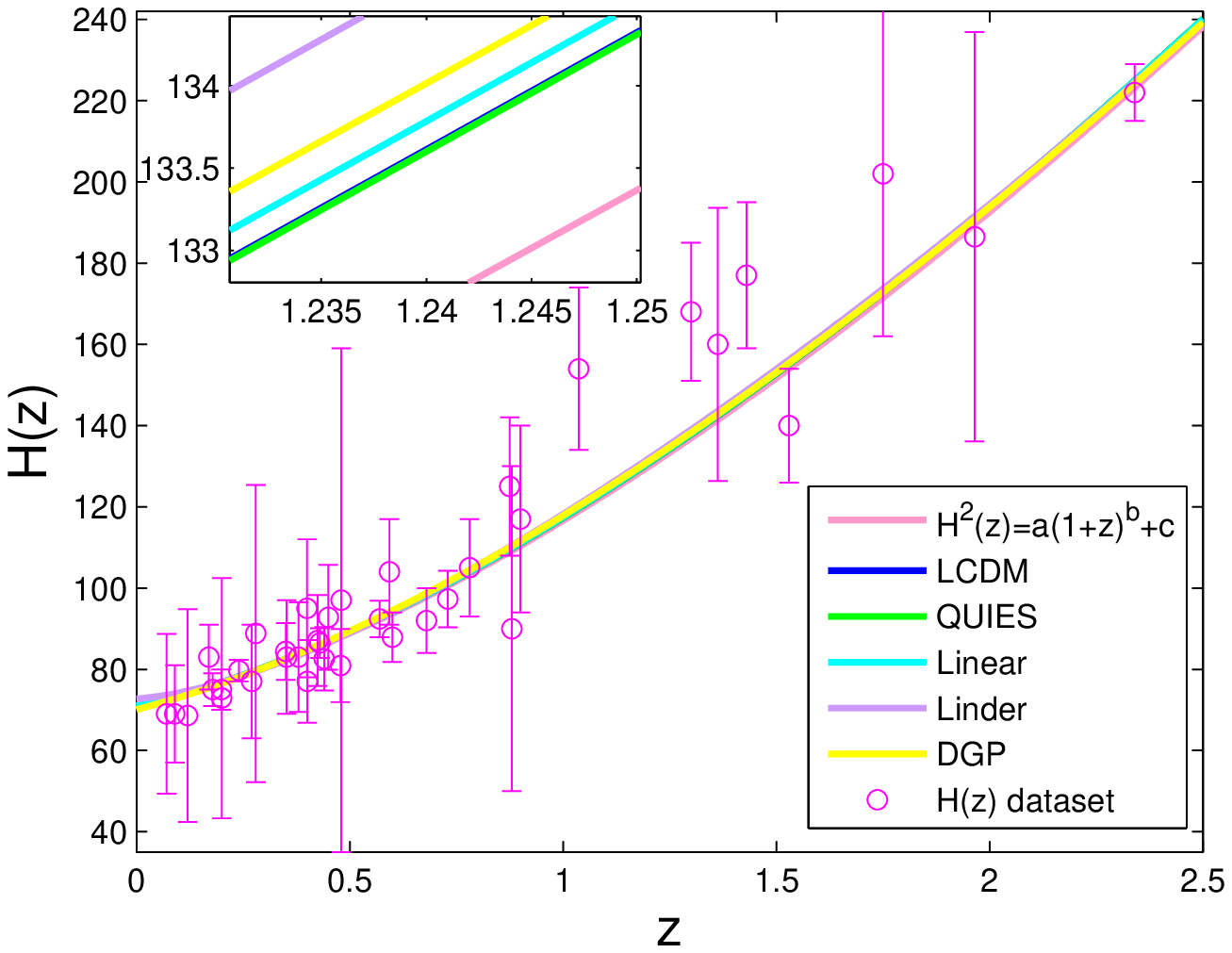}&
\includegraphics*[width=75.3mm,height=6cm]{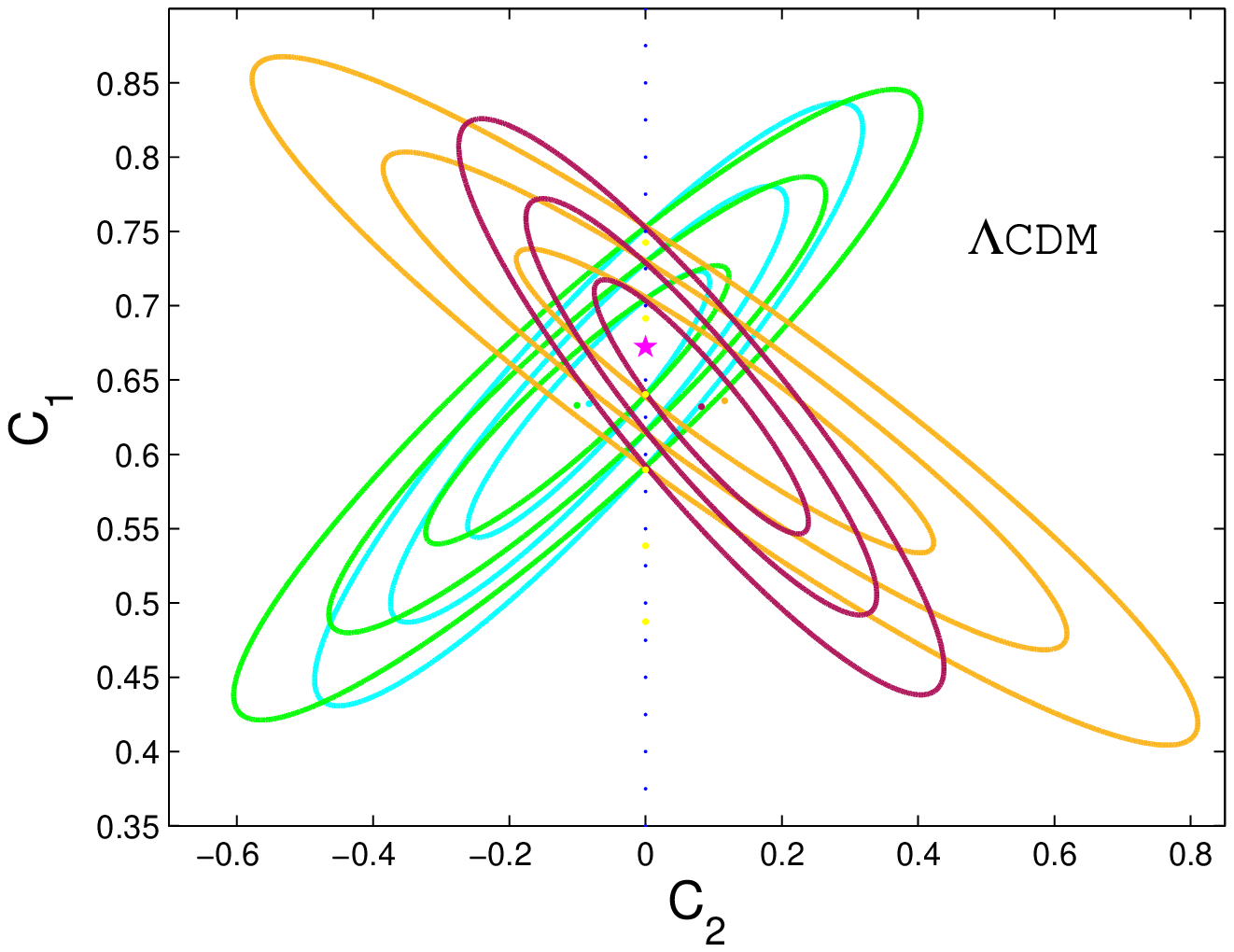}\\
\includegraphics*[width=75.3mm,height=6cm]{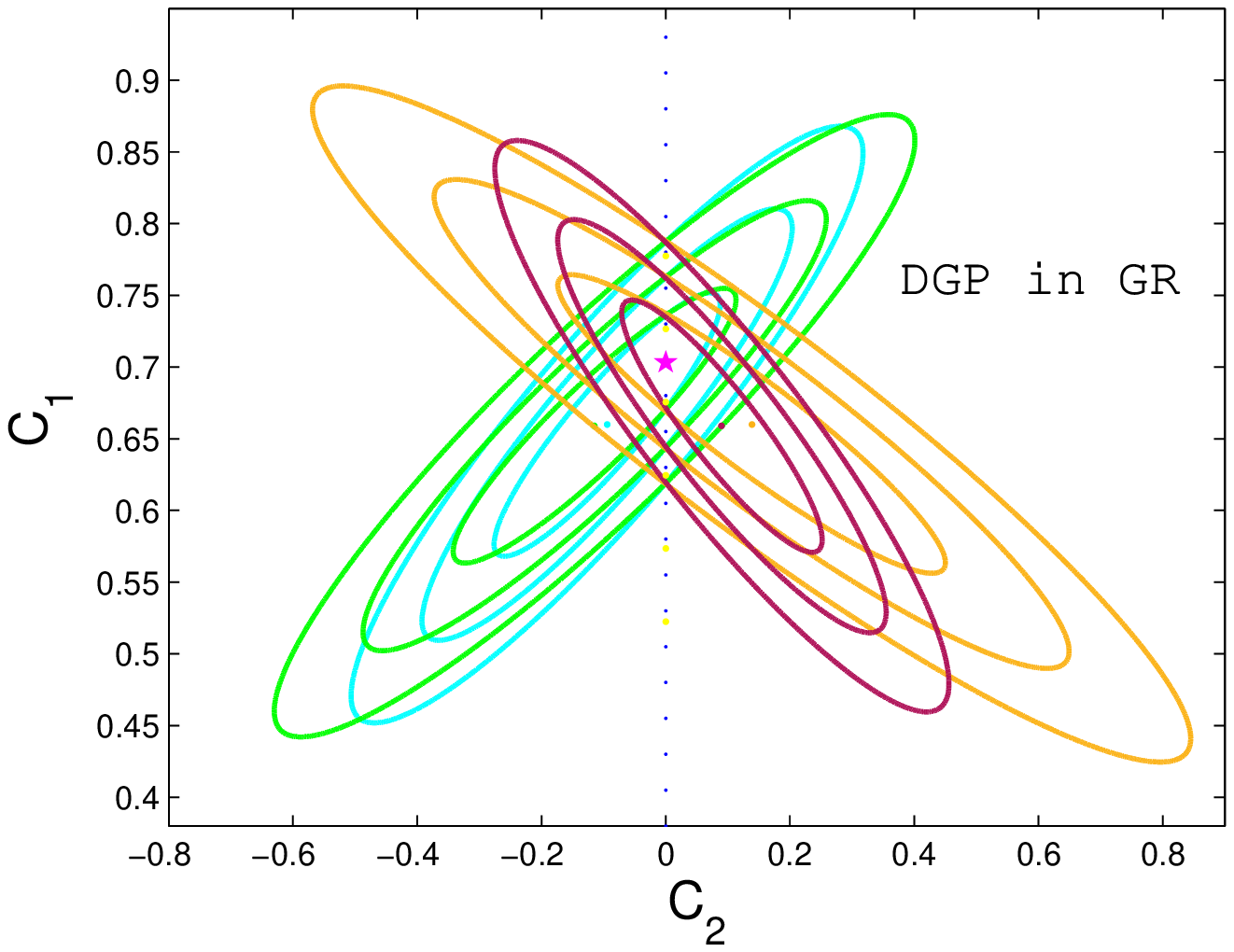}&
\includegraphics*[width=75.3mm,height=6cm]{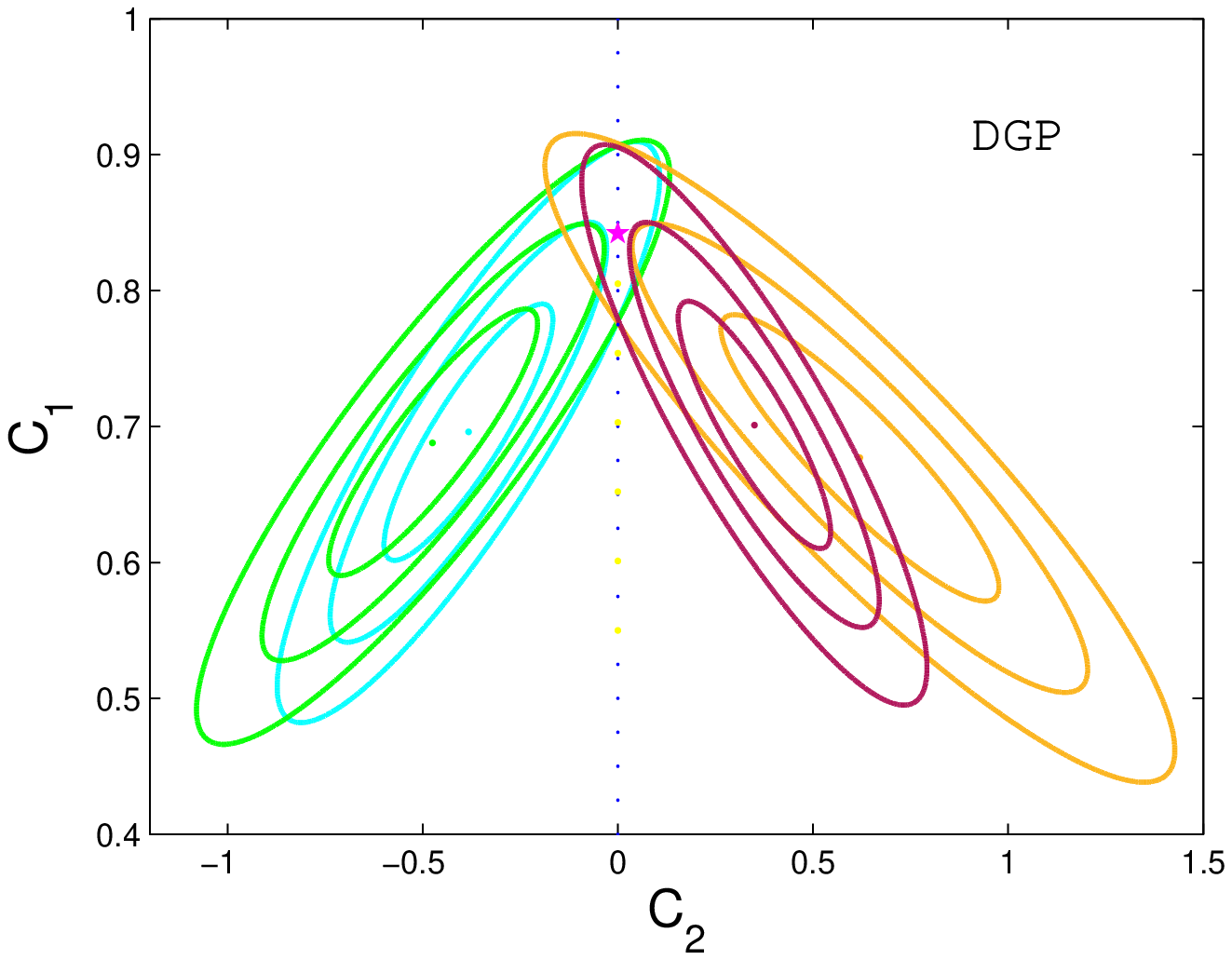}
\end{tabular}
\caption{Comparison of models and the testing results of $\Lambda$CDM, ``DGP in GR'' and DGP model. Displayed are the comparison of the $H(z)$ fitting traces from various cosmological models (top-left panel). The top-right panel shows the result of testing $\Lambda$CDM model using CRO. As for the bottom panels, there are the results of the ``DGP in GR''(bottom-left panel) and the modified version for DGP case(bottom-right panel). The description of the contours is as same as that in Figure 3. }
\end{center}
\end{figure*}

\section{Summary }

In this paper, we construct consistency relations between a kinematic probe, Hubble parameter $H(z)$, and a dynamical probe, the growth rate $R(z) (f\sigma_8(z))$ deduced from density fluctuation $\delta(z)$, and test their confidence level. The consistency relation should hold if general relativity is the correct theory of gravity in the universe, and the presence of significant deviations from the consistency can be used as the signature of a breakdown of general relativity at cosmological scales.

To quantify the relations between the kinematic and dynamical probes, we set up three consistency relations and test two of them in data experiment using parametric and non-parametric method. To model any departure from consistency relations, we employ two-dimensional parameterizations for a possible redshift dependence of the consistency relation to avoid making uncertain parameter $\Omega_{m0}\frac{\sigma_8(0)}{\delta(0)}$ involved in. Moreover, we propose trigonometric functions as efficient tools for parameterizations. As for both parametric method (Method A) and non-parametric method (Method B) for Hubble parameter, the theoretical results show us that in all two-dimensional parameterizations in the testing of CRO and CRI, there is no departure from the consistency relation in the $1\sigma$ region. In sum, the present observational Hubble parameter data and growth data favor reasonably that the general relativity is the correct theory of gravity on cosmological scales.

Furthermore, in order to confirm the validity of our test, we introduced a model of modified gravity, DGP model in it. It favors different gravitational theories from general relativity and the consistency relations based on the function of density perturbations would be changed. The fitting result of theoretical expression of $H(z)$ in DGP model with observational Hubble data has quite similar trace with that of $\Lambda$CDM model. Moreover, we compared the testing results of $\Lambda$CDM, ``DGP in GR'' and DGP model with CRO, which has shown tighter constraint in former test, and its modified version with fourth order Runge-Kutta method for DGP model. Finally we got that, $\Lambda$CDM and ``DGP in GR'' model passes the test with high consistency, while the result of DGP model obviously degrades. That is to say, based on the same expression of $H(z)$ which provides quite similar data with $\Lambda$CDM model, the change of testing principle leads to apparently disparate results. Considering that the fitting trace of $H(z)$ data in DGP model is quite similar to $\Lambda$CDM, the failure of its testing can't be blamed upon the inconsistency use of $H(z)$. It's due to the change of principle. It can be seen that it is the establishing of consistency relations which dominates the results of the testing.

Overall, according to the present observational Hubble data and growth rate data $f\sigma_8$, the general relativity is the correct theory of gravity on cosmological scales. It is desirable that the precise growth rate data in the high redshift region will refine the consistency relation between the kinematic and dynamical probes and we hope that our results would provide better directional information on testing general relativity.

It's indubitable that with the quality and quantity of cosmological observations
improving and unremitting efforts, the ultimate truth will be more and more clear. Thanks to all predecessors' marvelous work, of which the reverberations echo still.
\vskip 1cm


\acknowledgments

Thanks for all helpful comments and discussions. We appreciate the help so much with Dr. Xiao-Lei Meng, Dr. Shuo Cao and our research group for improvement of the paper. This work was supported by the National Science Foundation of China (Grants No. 11573006,11528306), the Ministry of Science and Technology National Basic Science program (project 973) under grant No. 2012CB821804, the Fundamental Research Funds for the Central Universities.


\end{document}